\documentclass[pdflatex,sn-nature,iicol]{sn-jnl}
\setcitestyle{super,comma,open={},close={}}

\usepackage{graphicx}
\usepackage{multirow}
\usepackage{amsmath,amssymb,amsfonts}
\usepackage{mathtools}
\usepackage{array}
\usepackage{xcolor}
\usepackage{float}
\usepackage{dsfont}
\usepackage{chemformula}
\usepackage[version=3]{mhchem}
\usepackage{subcaption}
\DeclareCaptionLabelSeparator{ncpipe}{ $|$ }
\newcommand{\Id}{\mathrm{1}\kern-.30em\mathrm{l}}

\newcommand*{\GW}{\textit{GW}}

\makeatletter
\def\ps@headings{\let\@mkboth\@gobbletwo\def\@oddhead{\hfill\thepage}\let\@evenhead\@oddhead\let\@oddfoot\@empty\let\@evenfoot\@empty}
\let\ps@titlepage\ps@headings
\makeatother

\begin{document}

\title{Learning compact and transferable representations of electron–hole interactions for large-scale excited-state simulations}

\author*[1]{\fnm{Christian} \sur{Venturella}}\email{christian.venturella@yale.edu}
\author[1]{\fnm{Daniel S.} \sur{King}}
\author[1]{\fnm{Ke} \sur{Liao}}
\author[1]{\fnm{Jiachen} \sur{Li}}
\author*[1]{\fnm{Tianyu} \sur{Zhu}}\email{tianyu.zhu@yale.edu}

\affil[1]{\orgdiv{Department of Chemistry}, \orgname{Yale University}, \orgaddress{\city{New Haven}, \state{CT}, \country{USA}}}

\abstract{Predicting electronic excitations in molecules and materials requires an accurate description of electron--hole interactions, whose computational cost grows steeply with system size. Here we present neural adaptive response functions (NARF), a machine learning framework that learns compact, atom-centered bases for electron--hole interactions, enabling excited-state calculations at much lower cost. By predicting these bases from local atomic environments, a model trained on only 706 small molecules reproduces the lowest singlet and triplet excitation energies with root-mean-square errors of 16--35~meV, while accelerating range-separated hybrid time-dependent density functional theory calculations by up to 100-fold. The same model transfers to larger systems and different density functionals without retraining, and to transition-metal complexes with fine-tuning. Furthermore, a model trained on mean-field data for small nanoclusters remains accurate in many-body \GW{} and Bethe--Salpeter calculations of unseen nanoclusters and periodic solids. This work opens up opportunities for efficient large-scale excited-state simulations, as illustrated by the absorption spectrum of a 2,520-atom pigment ring of a bacterial light-harvesting complex.}

\maketitle

\section{Introduction}
Computational prediction of electronic excitations in molecules and materials is central to applications in energy harvesting, imaging, sensing, and photoprotection~\cite{gonzalez2012progress,improta2016quantum,pyzerknapp2015highthroughput}. In large-scale molecular and materials simulations, linear-response time-dependent density functional theory (TDDFT) has been a popular choice for simulating excitation spectra~\cite{parr1989densityfunctional,perdew1996generalized,runge1984densityfunctional,laurent2013tddft,liang2022revisiting,chai2008longrange}.
For solid-state materials, many-body perturbation theory provides a more accurate description through the \GW{} approximation for charged quasiparticle excitations~\cite{golze2019gw,aryasetiawan1998gw,deslippe2012berkeleygw} and \GW{} combined with the Bethe–Salpeter equation (\GW+BSE) for neutral excitations~\cite{faber2014excited,blase2020bethe}.
Despite their different formalisms, these approaches all center on the treatment of particle--hole response, which determines neutral excitations in TDDFT and BSE and electronic screening in \GW{}. A shared computational challenge is thus to capture the interactions among a rapidly growing number of electron--hole transitions in a compact and accurate representation. In particular, describing spatially delocalized and charge-transfer excitations requires a sufficiently flexible orbital basis, which limits the scale of accessible excited-state simulations.
\par
The electron--hole interaction kernel is a two-particle quantity whose matrix representation grows quartically with system size. To reduce the associated cost, one common approach is the resolution of the identity (RI), also known as density fitting, which fits orbital-pair densities to an auxiliary basis to obtain compact factorizations of the underlying interaction integrals~\cite{ye2021fast,sun2017gaussian,lazarski2015density,kallay2014systematic,hellmann2022automated}. Standard auxiliary bases, however, are not optimized for electron--hole interaction kernels. Approaches with a fixed minimal auxiliary basis, such as TDDFT-as and TDDFT-ris, enable low-cost TDDFT simulations of molecular systems~\cite{giannone2020minimal,zhou2023tddftris}. Nevertheless, these approaches neglect the exchange--correlation kernel and rely partly on error cancellation, which limits their accuracy for triplet excitations and localized excitations in transition-metal complexes~\cite{zhou2023tddftris}. Tensor factorizations and low-rank representations have also gained popularity for many-body calculations of molecules and solids, such as tensor hypercontraction~\cite{parrish2013exact}, interpolative separable density fitting~\cite{yeh2024lowscaling}, and the projective dielectric eigenpotential method~\cite{govoni2015largescale,Yu2025WESTace,dong2021dielectric}. More recently, direct low-rank compression of the electron--hole interaction kernel has been used to accelerate time-dependent \GW{} calculations~\cite{Hou2025lowrankBSE}. Although systematically improvable, these approaches require constructing system-specific representations, introducing an upfront cost that grows with the size of the full basis. A remaining challenge is thus to combine response-adapted compression with the transferability of atom-centered basis functions across different systems and response theories.
\par
Machine learning (ML) offers a promising route to accurate and affordable materials modeling~\cite{butlerMachineLearningMolecular2018,deringerGaussianProcessRegression2021,keithCombiningMachineLearning2021,vonlilienfeldExploringChemicalCompound2020,kingCartesianEquivariantRepresentations2025,venturella2025unified,venturella2024machine,zhang2026transferable,schutt2018machine,zauchner2023accelerating,shao2023machine,lin2025deeplearning}, but most ML methods for electronic structure focus on the ground state. For excited states, ML models have been developed to predict singlet and triplet excitation energies directly from the atomic structure~\cite{westermayr2021excitedstates,dral2021lens,gomezbombarelli2016design,ramakrishnan2015spectra,ren2025sizetransferable,cignoni2023machine} and to fit excited-state potential energy surfaces~\cite{dral2018nonadiabatic,westermayr2019photodynamics,bai2022hohenberg,martyka2025charting,Unzueta2026mlnad,cignoni2024constrained}. However, learning transferable representations of electron--hole interactions across diverse molecular and material systems remains challenging for two reasons. First, these interactions contain long-range Coulomb and nonlocal exchange contributions that are difficult to infer from local atomic environments alone. Second, the excited-state Hamiltonian in the particle--hole basis grows rapidly with system size, making direct prediction impractical.
\par
Here we introduce neural adaptive response functions (NARF), an ML approach that learns compact, atom-centered orbital and auxiliary bases for electron--hole interactions across diverse systems and response theories. For each training system, we apply the interaction kernel to trial electron--hole amplitudes and project the result onto atom-centered functions to construct small matrices, which we call particle--hole covariates. In a new system, an equivariant neural network predicts each atom's covariates from its local environment. For the atomic orbital (AO) basis, basis functions are first selected to represent the ground-state density. One set of learned covariates  identifies additional functions needed to describe the occupied and virtual orbitals involved in electron--hole transitions. For the RI auxiliary basis, a second set of covariates identifies the functions needed to represent Coulomb and exchange interactions. The covariates also rank these basis functions by importance, so the basis can be systematically enlarged by lowering a single truncation threshold. We then perform standard TDDFT, \GW{}, or BSE calculations in the reduced bases. NARF thus gives access to most quantities the full calculation provides, such as excitation energies, oscillator strengths, and excited-state forces in TDDFT. In addition, we develop a complementary $\Delta$-ML correction model to further reduce excitation-energy errors due to basis truncation.
\par
Trained on 706 small chromophores, NARF retains only about 60\% of the AO basis and less than 20\% of the auxiliary basis, yet reproduces the singlet and triplet excitation energies of unseen molecules with root-mean-square errors of 16--35~meV and accelerates TDDFT by up to 100-fold, relative to full $\omega$B97X-D/def2-SVPD calculations with GPU4PySCF~\cite{li2025gpu4pyscf,wu2025gpu4pyscf}. Without retraining, the model transfers to much larger chromophores and to density functionals not used in training, and fine-tuning on only 208 complexes extends it to diverse transition metal compounds with similar accuracy. As a demonstration of scalability, we use NARF to calculate range-separated hybrid TDDFT absorption spectra of the B800 and B850 pigment rings of the bacterial light-harvesting complex 2 (LH2), with the larger 2,520-atom B850 calculation taking about twenty hours on four GPUs. Furthermore, a separate model trained on static mean-field kernels of small nanoclusters provides compact auxiliary bases for \GW{} and \GW{}+BSE calculations, reproducing full-basis band gaps and excitation energies to within about 30~meV in unseen nanoclusters and, without any periodic training data, to within 0.1~eV in bulk solids. These results suggest that the electron--hole response of small molecules and clusters already contains much of the local information needed to describe delocalized excitations in large molecules and molecular assemblies as well as many-body screening in solids.
\section{Results}
\subsection{Learning compact response bases}
\begin{figure*}[!t]
    \centering
\includegraphics[width=1\linewidth]{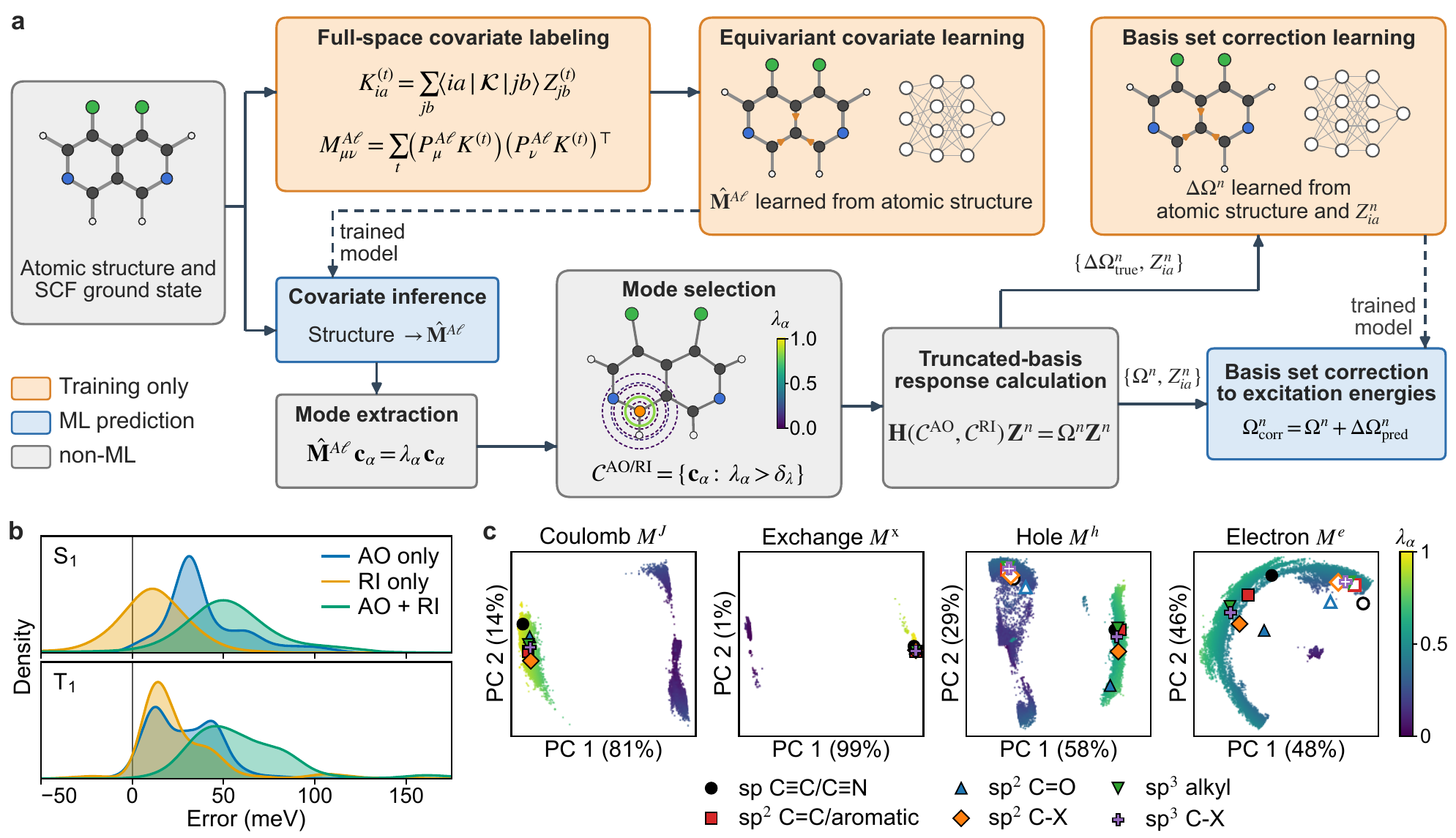}
\caption{\textbf{NARF workflow and the NARF covariate analysis.}
\textbf{a}~NARF workflow, starting from the atomic structure and the converged
SCF ground state. An equivariant neural network
predicts covariates from the atomic structure. Their leading modes define
the AO and RI auxiliary bases for the response calculation, and a second
network corrects the resulting excitation energies. In the mode-selection box,
rings show the eight RI $s$ modes predicted on one carbon, with radius
giving the spatial extent and color the weight $\lambda_{\alpha}$; only the
leading mode (solid) passes the threshold.
\textbf{b}~Distributions of signed $\mathrm{S}_1$ and $\mathrm{T}_1$ errors from
truncated-basis TDA over the 706 NARF-mol training molecules, with the AO basis,
the RI basis or both truncated using the true covariates ($\delta_{\lambda} = 0.2$,
no $\Delta\Omega$ correction), against full-basis TDA at the $\omega$B97X-D/def2-SVPD level.
$\mathrm{S}_1$/$\mathrm{T}_1$ RMSEs are 45/33, 35/37 and
62/69~meV, respectively.
\textbf{c}~Radial $s$ functions of all 9,611 carbon atoms in the NARF-mol training
set for the RI-basis (Coulomb, exchange) and AO-basis (hole,
electron) covariates, weighted by $\sqrt{\lambda_{\alpha}}$, projected onto
the first two principal components of each covariate (percentage of variance
captured by each component shown on the axes) and colored by
$\lambda_{\alpha}$. Markers give the mean position of the leading mode
(filled), and for $M^{h}$ and $M^{e}$, the second mode (open) in each
carbon environment. X denotes a heteroatom.}
\label{fig:narf_scheme}
\end{figure*}

To cast electron--hole interactions into a representation suitable for learning, we introduce the central object of NARF: a particle--hole covariate \(M^{A\ell}\) for each atom \(A\) and angular momentum shell \(\ell\). The covariates take the general form (Fig.~\ref{fig:narf_scheme}a)
\begin{subequations}
  \begin{align}
    M^{A\ell}_{\mu\nu} &= \sum_{t} \left(P^{A\ell}_{\mu} K^{(t)} \right) \left( P^{A\ell}_{\nu} K^{(t)} \right)^{\top}\\
  K^{(t)}_{ia} &= \sum_{jb} \langle ia \mid \mathcal{K} \mid jb \rangle Z^{(t)}_{jb}.
  \end{align}
  \end{subequations}
Here $K^{(t)}_{ia}$ is the action of the full-space response kernel $\mathcal{K}$ on a trial electron--hole amplitude vector $Z^{(t)}$, where $i,j$ denote occupied orbitals, $a,b$ denote virtual orbitals, and $t$ indexes the trial vectors, which can span all electron--hole transitions or a subspace of interest (e.g., valence excitations). The operator $P^{A\ell}_{\mu}$ maps these actions onto the radial functions $\mu,\nu$ of a parent orbital or auxiliary basis for atom $A$ and shell \(\ell\), and its exact form depends on the basis being compacted and on the interaction channel. Each covariate $M^{A\ell}$ can be interpreted as a covariance that measures how much response-kernel signal maps onto the parent functions of one atomic shell, and it is positive semidefinite by construction.

\par In the training stage, reference systems are labeled with two sets of covariates. For the AO basis, $M^{0}$ is obtained by direct projection of the ground-state electron density onto the parent basis and need not be learned, while $M^{h}$ and $M^{e}$
capture the hole and electron channels of the response, i.e., the occupied and virtual orbitals involved in the transitions. For the RI auxiliary basis, $M^{J}$ and $M^{\text{x}}$ capture the Coulomb and exchange channels. The channel-specific definitions are given in the Methods (Eqs.~\ref{eq:cov_ground}--\ref{eq:covx}). The labels are then orthogonalized and trace-normalized, giving $\hat{M}^{A\ell}$, which removes their arbitrary dependence on energy scales and local integral metrics. Finally, an E(3)-equivariant neural network (i.e., GotenNet~\cite{aykent2025gotennet}) learns the four
normalized covariates $\hat{M}^{A\ell}$ of each atomic shell from the atomic structure, with an output layer that keeps the predicted covariates positive semidefinite.

\begin{figure*}[!t]
    \centering
\includegraphics[width=1\linewidth]{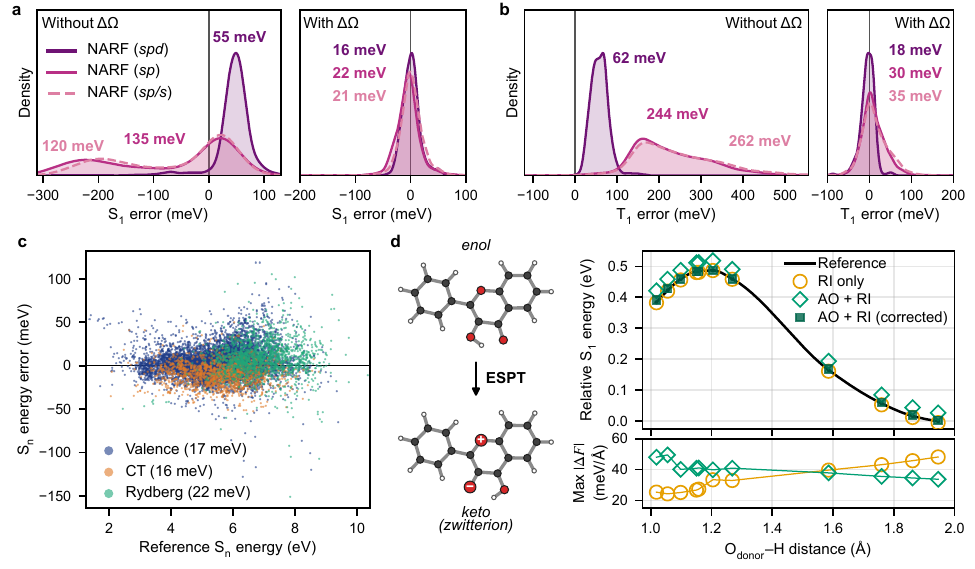}
\caption{\textbf{Accuracy of NARF-mol on unseen molecules.}
All errors are for truncated-basis TDA with NARF-mol against full-basis TDA
at the $\omega$B97X-D/def2-SVPD level.
\textbf{a}, \textbf{b} Distributions of signed errors in the lowest singlet
and triplet excitation energies of the 992 test molecules, for the NARF
\textit{spd}, \textit{sp} and \textit{sp/s} RI settings with the same AO
truncation, without and with the $\Delta\Omega$ correction. Numbers give the
RMSE of each distribution. The correction is applied to each root before
re-sorting, and the \textit{sp} and \textit{sp/s} settings use
$\Delta\Omega$ models fine-tuned for their auxiliary spaces.
\textbf{c} Errors of the ten lowest singlet states of the test molecules
with the \textit{spd} setting and the $\Delta\Omega$ correction, colored by
the character of the reference state: valence, charge transfer (CT) or
Rydberg (RMSEs in the legend; see state assignment criteria in Methods).
\textbf{d} Excited-state proton transfer (ESPT) in 3-hydroxyflavone, from
the enol to the zwitterionic keto form. Top: S$_1$ energy relative to the
path minimum, from the full basis (reference) and from NARF (\textit{spd}) with
RI-only truncation, AO + RI truncation, and AO + RI truncation with the
$\Delta\Omega$ correction. Bottom: maximum atomic force error along the
path. O$_{\mathrm{donor}}$ is the proton-donating oxygen.}
\label{fig:small_molecules_results}
\end{figure*}

\par
In the inference stage (Fig.~\ref{fig:narf_scheme}a), NARF first predicts the normalized covariates for unseen systems from the atomic structure. The predicted covariates for each $A\ell$ are then diagonalized to obtain basis functions (modes) $\mathbf{c}_\alpha$, each with an eigenvalue $\lambda_{\alpha}$ between 0 and 1 that gives the fraction of the particle-hole response it carries
\begin{subequations}
  \label{eq:cov_eigval}
  \begin{gather}
  \hat{\mathbf{M}}^{A\ell}\,\mathbf{c}_\alpha=\lambda_\alpha\,\mathbf{c}_\alpha\\
  \mathcal{C}^{\mathrm{AO/RI}} = \{\mathbf{c}_\alpha:\hat{\mathbf{M}}^{A\ell}\mathbf{c}_\alpha=\lambda_\alpha\mathbf{c}_\alpha,\,\lambda_\alpha>\delta_\lambda\}
\end{gather}
\end{subequations}
Basis functions with $\lambda_\alpha$ above the user-defined threshold $\delta_{\lambda}$ are retained, so $\delta_{\lambda}$ controls the cost accuracy tradeoff of the truncated bases. For most of our tests, we use a single threshold $\delta_\lambda$, though in select benchmarks, we instead retain the leading modes up to a cumulative weight $\theta_{\lambda}$: $\sum_\alpha \lambda_\alpha < \theta_\lambda$.

\iffalse
After covariate inference and basis function selection, a standard response theory calculation proceeds in the truncated AO and RI bases:
\begin{subequations}
\label{eq:truncation}
\begin{gather}
(pq|rs) \approx \textstyle\sum_{PQ\in\mathcal{C}^{\mathrm{RI}}}(pq|P)\,[\mathbf{V}^{-1}]_{PQ}\,(Q|rs)\\
\phi_p \approx \textstyle\sum_{\mu\in\mathcal{C}^{\mathrm{AO}}}C_{\mu p}\,\chi_\mu\\
 \mathbf{H}(\mathcal{C}^{\mathrm{AO}},\mathcal{C}^{\mathrm{RI}})\vert\Phi^n\rangle=\Omega^n\vert\Phi^n\rangle
\end{gather}
\end{subequations}
where $V_{PQ}=(P|Q)$ is the Coulomb metric of the retained auxiliary functions and $C_{\mu p}$ are the molecular-orbital coefficients expanded in the truncated AO space via projection.
\fi

A standard response calculation, such as TDDFT within the Tamm--Dancoff approximation (TDA), then starts from a converged self-consistent-field (SCF) ground state and proceeds in the truncated bases:
\begin{equation}
\label{eq:truncation}
  \mathbf{H}\bigl(\mathcal{C}^{\mathrm{AO}}, \mathcal{C}^{\mathrm{RI}}\bigr)
    \,\mathbf{Z}^{n} = \Omega^{n}\,\mathbf{Z}^{n},
\end{equation}
where the response Hamiltonian $\mathbf{H}$ is built from density-fitted integrals in the truncated bases, with the molecular orbitals projected onto the retained AO functions, and is diagonalized to obtain the excitation energies $\Omega^{n}$ and amplitudes $\mathbf{Z}^{n}$. Because the covariates are predicted from the atomic structure, a new calculation never evaluates the response in the full parent basis. Moreover, shared local parameters permit a model trained on small systems to be evaluated on much larger ones, while the local functions can still describe nonlocal excitations via their interaction in the global $\mathbf{H}$. We note that NARF is also easily compatible with existing quantum chemistry methods and codes.

\par
While NARF typically yields excitation energies $\Omega^n$ with errors of 0.1 eV or smaller, higher accuracy can be achieved by training a complementary $\Delta$-ML basis set correction model (Fig.~\ref{fig:narf_scheme}a). For each excited state $n$, we define local features that resolve the electron density and Coulomb potential missed by a given basis truncation. Then we train a second equivariant neural network to predict the energy error $\Delta\Omega^n$ as a learned linear weighting of these features. Details of the derivations and ML architectures are given in the Methods.
\iffalse
\begin{subequations}
\begin{align}
\mathbf{w}_{A} &= D^\mathbf{w}(h^\mathbf{w}_A(\{Z_B,\mathbf{r}_B\}_{B\in d}))\\
\Delta\Omega^n &=  \sum_{A}(\mathbf{w}_{A})^T\mathbf{g}_{n,A}
\end{align}
\end{subequations}
\fi

To assess the proposed basis truncation independently of ML, we studied 706
small-molecule chromophores from the literature
\cite{Stuke2020OE62,LupoPasini2023ExcitedStates,Joung2020ChromDB,StJohn2019MPNN,Omar2022Repurposing,Kim2025PubChem,Gupta2021BODIPY,Kim2023OMG},
covering H, B, C, N, O, F, Si, P, S, Cl, As, Se, Br, Mg, Zn, Zr and Cd,
at the $\omega$B97X-D/def2-SVPD level
\cite{chai2008longrange,weigend2005balanced,rappoport2010property}. Starting from the converged
ground-state orbitals, we computed the covariates and full-basis TDA
reference energies using our GPU-accelerated implementation built on
PySCF~\cite{sun2020recent,sun2026pyscf}.
Because the triplet channel of the response kernel lacks the
Coulomb term, the singlet and triplet bases are selected separately. On
average, truncation shrinks the AO basis by $1.6\times$ and the RI basis by
$5\times$ or more. With these computed (true) covariates, TDA with AO-only or
RI-only truncation gives $\mathrm{S}_1$/$\mathrm{T}_1$ RMSEs of 45/33 or
35/37~meV, while TDA with the combined truncation gives 62/69~meV, against
full $\omega$B97X-D/def2-SVPD TDA calculations
(Fig.~\ref{fig:narf_scheme}b). These errors are predominantly blue shifts,
which motivates us to develop the $\Delta\Omega$ correction scheme.

\begin{figure*}[!t]
    \centering
\includegraphics[width=\linewidth]{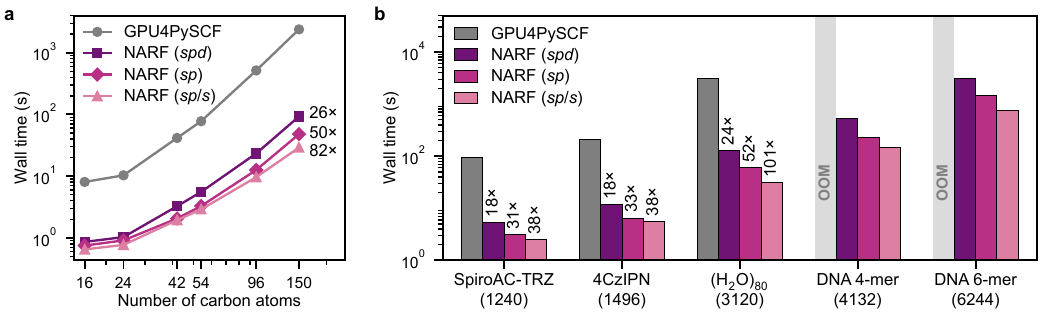}
\caption{\textbf{Speedup of molecular TDA calculations with NARF.}
Wall times for 10-root singlet $\omega$B97X-D/def2-SVPD TDA
calculations on one H200 GPU using GPU4PySCF~\cite{li2025gpu4pyscf,wu2025gpu4pyscf} with density fitting in the full
basis and NARF-mol with the \textit{spd}, \textit{sp} and \textit{sp/s}
RI settings with the same AO truncation.
\textbf{a} Hydrogen-terminated graphene flakes from C$_{16}$H$_{10}$ to
C$_{150}$H$_{30}$. Speedups are shown for the largest flake.
\textbf{b} Chemically diverse systems. Parentheses give the number of AO
basis functions in the full def2-SVPD basis. OOM indicates that GPU4PySCF exhausted the
available 1.9\,TB of host memory while building the three-index
density-fitting integrals.}
\label{fig:timing}
\end{figure*}

To understand which basis functions are retained after truncation, we mapped the
radial $s$ functions of all 9,611 carbon atoms in this small-molecular dataset
by their importance-weighted shapes
(Fig.~\ref{fig:narf_scheme}c). Both RI covariates, $M^J$ and
$M^{\text{x}}$, are effectively rank one: each carbon keeps one $s$ function
carrying over 80\% of the weight, and this function has the same shape in all
six bonding environments. For the AO covariates, $M^h$ needs on average two
functions per carbon, again with little variation between environments. By
contrast, the two leading $M^e$ functions are nearly equally important, and
which one dominates depends on bonding. For carbon, the RI and hole functions are
therefore largely fixed across chemical environments, while the electron
functions carry most of the environment dependence.

\subsection{Accuracy and cost of molecular excitation calculations}

\par
We trained NARF on these 706 molecules, naming the model NARF-mol, and
tested it on 992 held-out molecules with the same elements (on average 20
and at most 25 non-hydrogen atoms). NARF-mol predicts the covariates, and
the excited states are computed using TDA with the $\omega$B97X-D functional
in the predicted AO and auxiliary bases, with the same thresholds as in
Fig.~\ref{fig:narf_scheme}b. The AO truncation follows one rule throughout, while the auxiliary
truncation can be tuned to trade accuracy for cost. We use three
auxiliary-basis settings, named after the auxiliary shells they retain.
NARF (\textit{spd}) uses the $s$, $p$ and $d$ functions selected from the
covariates, with one fixed $s$ function on each hydrogen. NARF
(\textit{sp}) removes the hydrogen auxiliaries and the $d$ shells of
non-metal atoms. NARF (\textit{sp/s}) uses the same auxiliary space as NARF
(\textit{sp}) for the Coulomb and exchange--correlation kernel
($f_\mathrm{xc}$) terms and only $s$ functions on non-metal atoms for
exchange. Averaged over the 992 test molecules, NARF keeps 63\% of the
def2-SVPD AO basis in all three settings, and 19\%, 8\% and 8\% of the
def2-universal-jkfit auxiliary functions with NARF (\textit{spd}), NARF
(\textit{sp}) and NARF (\textit{sp/s}), respectively; with NARF
(\textit{sp/s}), only 2\% are kept for exchange. Each application below uses the setting suited
to its purpose, and we state the retained fractions of the AO and auxiliary
bases with each result.

With NARF (\textit{spd}), both the S$_1$ and T$_1$ energies exhibit a nearly
uniform blue shift (Fig.~\ref{fig:small_molecules_results}a,b), with RMSEs of 55 and
62~meV. These errors are comparable to those obtained with the true
covariates on the training set (62 and 69~meV,
Fig.~\ref{fig:narf_scheme}b), so nearly all of the error comes from the
basis truncation rather than from the covariate ML predictions (see Supplementary
Note 7). The $\Delta$-ML correction reduces the RMSEs to 16 and 18~meV,
within chemical accuracy (i.e., 1 kcal/mol, or 43 meV) across a wide range of excitation energies.
For NARF (\textit{sp}), the $\Delta\Omega$ correction fine-tuned for this
auxiliary space reduces the S$_1$/T$_1$ RMSEs from 135/244 to 22/30~meV.
For NARF (\textit{sp/s}), the corresponding RMSEs fall from 120/262 to
21/35~meV, so restricting exchange to $s$ functions has little effect on the
corrected S$_1$/T$_1$ RMSEs relative to NARF (\textit{sp}).
For NARF (\textit{spd}), we also categorized the lowest ten singlets across all
992 molecules (9,920 excited states) by their excitation character:
Rydberg, charge-transfer, or valence (Fig.~\ref{fig:small_molecules_results}c). The
corresponding error distribution shows no definitive preference for state
character or S$_n$ energy, emphasizing the balanced nature of the learned
basis functions.

\begin{figure*}[!t]
    \centering
\includegraphics[width=1\linewidth]{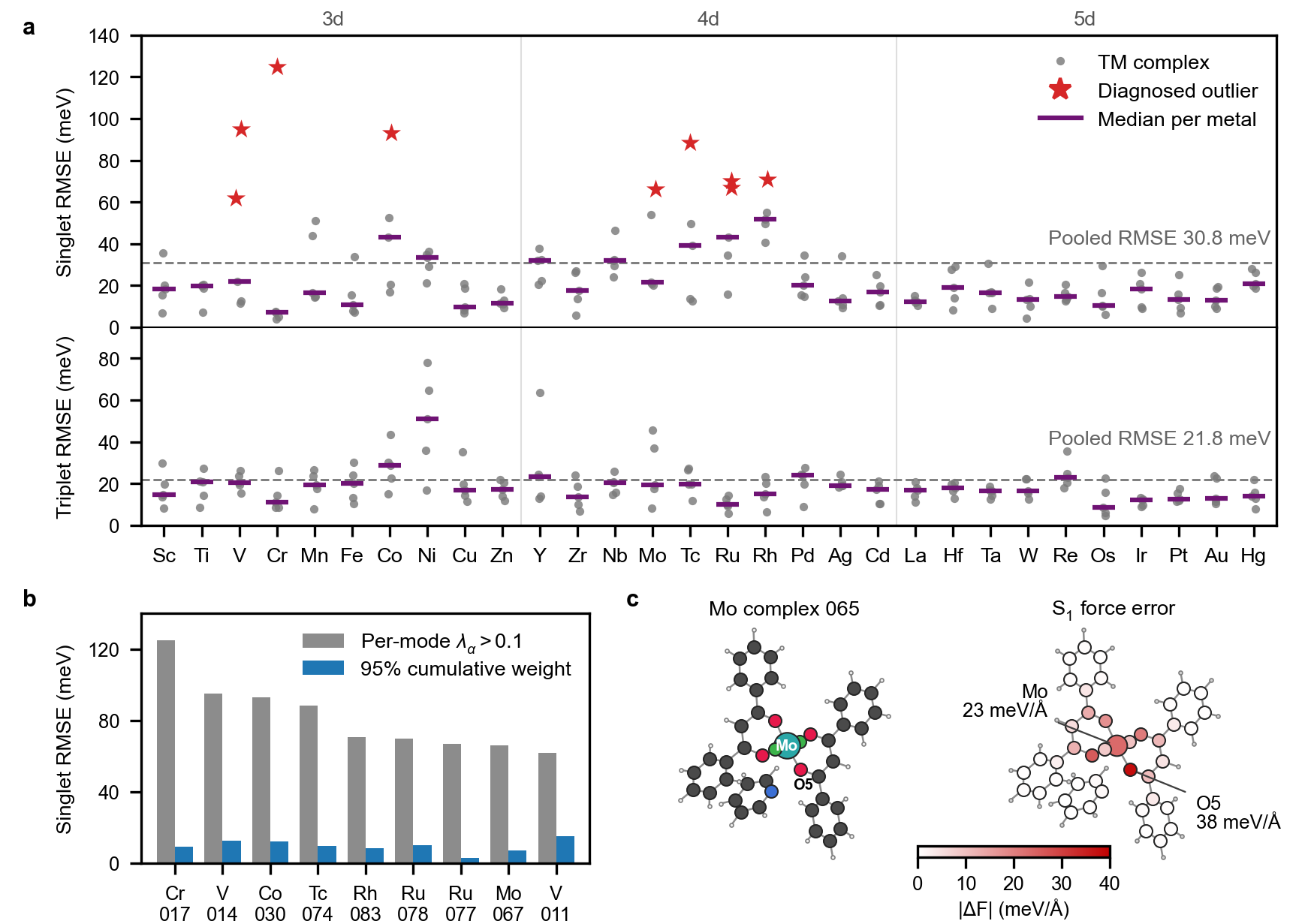}
\caption{\textbf{NARF fine-tuned for transition-metal complexes.}
Results for NARF-mol-tm on 150 held-out tmQM complexes, five per metal
($\omega$B97X-D/def2-SVPD TDA, ten roots per spin).
\textbf{a} RMSE of the ten lowest singlet (top) and triplet (bottom)
excitation energies of each complex, with the \textit{spd} setting (full AO
basis on the metal) and the $\Delta\Omega$ correction.
Purple bars mark the median for each metal, dashed lines the RMSE pooled
over all states, and red stars the nine singlet outliers examined in \textbf{b}.
\textbf{b} Singlet RMSE of the nine outliers with the default per-mode
threshold ($\lambda_{\alpha}>0.1$) and with a rule that retains the
leading auxiliary modes until their cumulative weight reaches 95\%.
Both rules apply to the metal and its coordinating atoms. Labels give
the metal and the complex index.
\textbf{c} Mo complex 065 and its per-atom
$\mathrm{S}_1$ force errors with the 95\% rule and no $\Delta\Omega$
correction (right). The errors on Mo and on the oxygen with the largest
error (O5) are labeled.}
\label{fig:tm}
\end{figure*}

\begin{figure*}[!t]
    \centering
\includegraphics[width=1\linewidth]{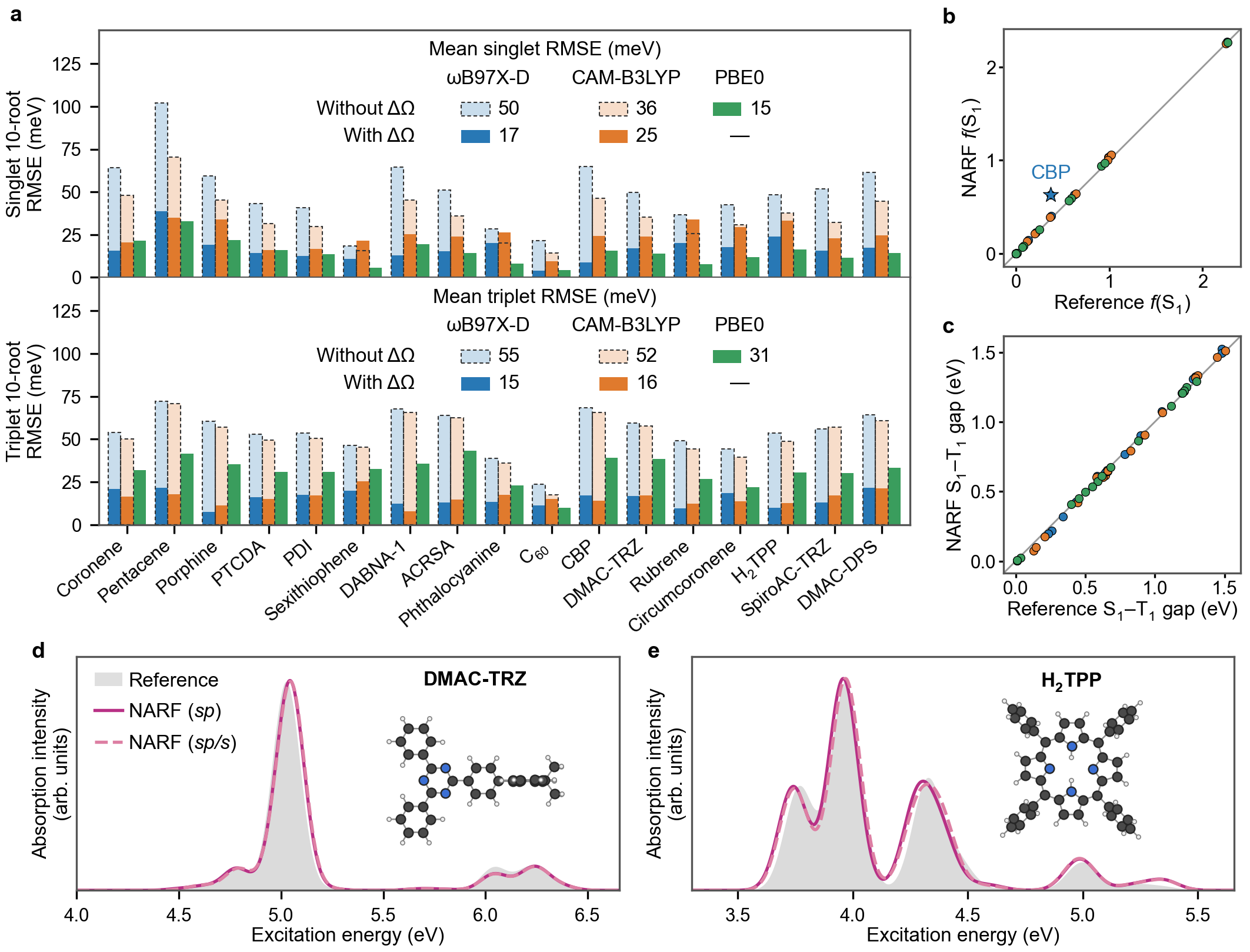}
\caption{\textbf{Transfer of NARF-mol across system sizes and density functionals.}
All excitation energies are computed using TDA, with full-basis def2-SVPD
calculations as the reference. The NARF-mol model was trained on
$\omega$B97X-D data only; CAM-B3LYP and PBE0 were not used in training.
\textbf{a} RMSE over the ten lowest singlet (top) and triplet (bottom)
states of 17 organic chromophores with the NARF ($spd$) setting, for the
three density functionals. For $\omega$B97X-D and CAM-B3LYP, light bars with
dashed outlines show RMSEs before the $\Delta\Omega$ correction and solid bars
show RMSEs after correction. The green PBE0 bars show uncorrected RMSEs.
The legends give the means over the
17 molecules.
\textbf{b}, \textbf{c} S$_1$ oscillator strength $f$(S$_1$) (\textbf{b}) and
S$_1$--T$_1$ gap (\textbf{c}) from NARF-mol against the full def2-SVPD
reference, for the same molecules and functionals (colors as in \textbf{a};
with $\Delta\Omega$ for $\omega$B97X-D and CAM-B3LYP). RMSEs for
$\omega$B97X-D, CAM-B3LYP and PBE0 are 0.065, 0.016 and 0.008 (\textbf{b})
and 25, 26 and 11~meV (\textbf{c}).
\textbf{d}, \textbf{e} $\omega$B97X-D absorption spectra from the 30 lowest
singlet excitations of DMAC-TRZ
(\textbf{d}) and H$_2$TPP (\textbf{e}) from NARF-mol with the \textit{sp} and
\textit{sp/s} settings, starting from a def2-SVP ground state with a
one-shot def2-SVPD update of the virtual orbitals, against the full
def2-SVPD reference. The NARF spectra include the $\Delta\Omega$ correction.}
\label{fig:transfer_chromophores}
\end{figure*}

\par
NARF bases also give accurate excited-state forces, without any training on
forces or excited-state dynamics trajectories. For the excited-state proton transfer of
3-hydroxyflavone~\cite{Casadesus2006flavone,Chevalier2012flavone}, from the enol form stable in S$_0$ to the keto
(zwitterionic) form preferred in S$_1$, we sampled structures along the
S$_1$ path and computed energies and forces with NARF-mol in the (\textit{spd}) setting without fine-tuning, using
the basis predicted at each structure (Fig.~\ref{fig:small_molecules_results}d). On
average, 67\% of the AO basis and 19\% of the auxiliary basis are kept.
Without the $\Delta\Omega$ correction in the gradient, the maximum atomic
force error stays below 50~meV~\AA$^{-1}$ along the whole path for both the
RI-only and the AO + RI truncation, and the $\Delta\Omega$-corrected
relative S$_1$ energies lie within 10~meV of the reference. The NARF basis
functions are thus transferable across geometries and give smooth
excited-state surfaces.
\par
Lastly, we compare the cost of NARF with density-fitted TDA in the full
basis using GPU4PySCF, for a 10-root TDA solve on one H200 GPU
(Fig.~\ref{fig:timing}). Across the systems shown, NARF (\textit{spd}),
NARF (\textit{sp}) and NARF (\textit{sp/s}) keep on average 61\% of the AO
basis and 16\%, 7\% and 7\% of the auxiliary basis, and all three are
consistently much faster than the full-basis calculation. For exchange, NARF
(\textit{sp/s}) keeps only about 2\% of the auxiliary basis, against 7\% in
NARF (\textit{sp}), which explains its additional speedup. On hydrogen-terminated graphene flakes from
C$_{16}$H$_{10}$ to C$_{150}$H$_{30}$ (Fig.~\ref{fig:timing}a), the speedups
are 9--26$\times$ for NARF (\textit{spd}) and 12--82$\times$ for NARF
(\textit{sp/s}), and they grow with system size. For SpiroAC-TRZ, 4CzIPN and
(H$_2$O)$_{80}$ (Fig.~\ref{fig:timing}b), the speedups range from
18$\times$ with NARF (\textit{spd}) to 101$\times$ with NARF
(\textit{sp/s}). For the DNA 4- and 6-mers, GPU4PySCF exhausts the 1.9~TB of
host memory, while NARF finishes in 2.4--53~min, depending on the system and
setting.

\subsection{Fine-tuning for transition-metal complexes}
Transition-metal (TM) complexes are a demanding test for basis reduction:
the metal $d$ manifold gives dense sets of low-lying metal-centered,
ligand-centered and charge-transfer states, and the compact pair densities
around the metal need auxiliary functions that organic chromophores rarely
use. To extend our model to these systems, we fine-tuned NARF-mol on 208
complexes from the tmQM database~\cite{Balcells2020tmQM}, covering 30 transition metals
from Sc to Hg, naming the model NARF-mol-tm. We also extended the auxiliary
functions on the metal to $s$--$g$ shells and retrained the RI part of the
$\Delta\Omega$ correction on the same data. The metal and its coordinating
atoms (non-hydrogen atoms within 1.3$\times$ the sum of covalent radii) are
treated more conservatively: the metal keeps its full def2-SVPD AO basis,
and on all of these atoms we keep every auxiliary mode with
$\lambda_{\alpha} > 0.1$ (default 0.2) in either the Coulomb or the exchange
covariate. All other atoms use the NARF (\textit{spd}) setting.\par
On a held-out set of 150 TM complexes (five per metal), the predicted bases
retain about 60\% of the AO basis and 20\% of the auxiliary basis, and the
$\Delta\Omega$-corrected RMSE pooled over the ten lowest states of all
complexes is 31~meV for singlets and 22~meV for triplets
(Fig.~\ref{fig:tm}a). For 19 of the 30 metals the median singlet RMSE is
below 20~meV, comparable to organic chromophores. The largest errors occur
for V, Cr, Co, Mo, Tc, Ru and Rh complexes with low-lying metal-centered and
metal-to-ligand charge-transfer states (stars in Fig.~\ref{fig:tm}a;
structures in Fig.~S2). These states need more radial
flexibility around the metal, and the covariates offer a direct route to it.
For these nine complexes, we keep the leading auxiliary modes on the metal
and its coordinating atoms up to a cumulative weight of 95\% (Fig.~S1).
This adds at most 12\% of the metal's parent RI basis and less
than 2\% of that of the full system, and it reduces their singlet RMSEs from
62--125~meV to 3--15~meV (Fig.~\ref{fig:tm}b).\par
The same rule also improves excited-state forces. For a Mo complex of 70 atoms, the S$_1$ force error on Mo without the $\Delta\Omega$
correction drops from 89~meV~\AA$^{-1}$ with the default threshold to
23~meV~\AA$^{-1}$ with the 95\% rule, and the largest error on any atom is
then 38~meV~\AA$^{-1}$, on a chelating oxygen (Fig.~\ref{fig:tm}c).\par

\subsection{Transfer across system sizes and density functionals}
We next return to the NARF-mol model, and apply it without
retraining to 17 large chromophores for organic electronics and energy
applications~\cite{Bannwarth2019GFN2,Knysh2024QUESTChrom,TchapetNjafa2026Screening}. Although most are larger and more extensively
conjugated than the training molecules, the truncation errors remain small.
With NARF (\textit{spd}), the model keeps on average 60\% of the AO basis
and 16\% (singlets) or 22\% (triplets) of the auxiliary basis. Over ten
singlet roots, the mean RMSE from truncation alone is 50~meV, and the
$\Delta\Omega$ correction reduces it to 17~meV (Fig.~\ref{fig:transfer_chromophores}a,
top). The triplets follow a similar pattern, with mean RMSEs of 55 and
15~meV (Fig.~\ref{fig:transfer_chromophores}a, bottom).

\begin{figure*}[!t]
    \centering
\includegraphics[width=\linewidth]{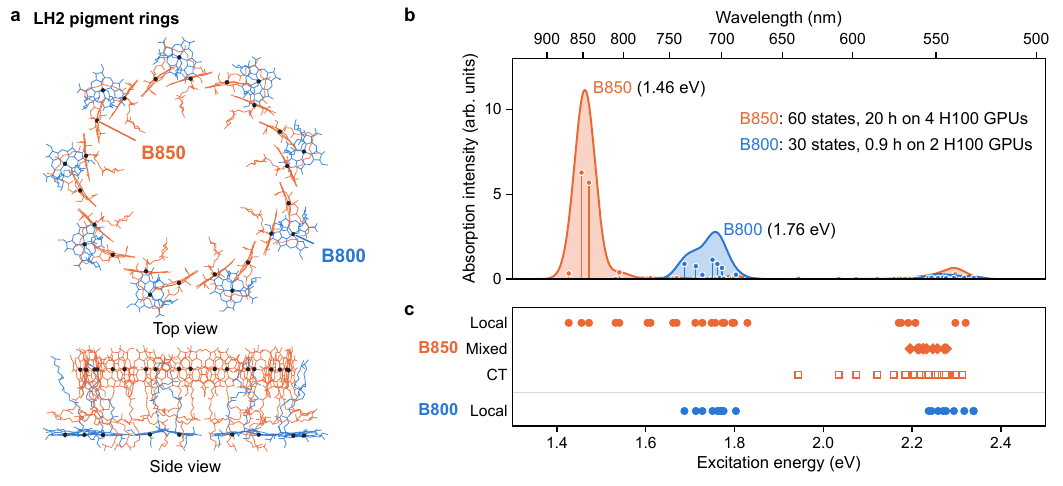}
\caption{\textbf{NARF-accelerated absorption spectra of the LH2 pigment
rings.}
\textbf{a} The 27 bacteriochlorophyll \textit{a} (BChl \textit{a}) pigments
of LH2 from \textit{Rhodoblastus acidophilus} (Protein Data Bank entry
2FKW), in top and side view: the B850 ring (orange, 18 BChl \textit{a}) and
the B800 ring (blue, 9 BChl \textit{a}). Black dots mark the Mg atoms.
\textbf{b} $\omega$B97X-D TDA absorption spectra of the two rings, computed
separately with NARF-mol in the \textit{sp/s} setting with $\Delta\Omega$
excitation energy correction, from a def2-SVP
ground state with a one-shot update of the virtual orbitals using def2-SVPD
on the ring and carbonyl atoms.
Curves are broadened with Gaussians (FWHM 0.05~eV), and sticks mark the
individual states by oscillator strength. The reported times are for the TDA solves.
\textbf{c} Character of the singlet states in the 1.3--2.5~eV window from a transition-density analysis,
classified by its charge-transfer (CT) weight: local ($<0.25$, filled
circles), mixed (0.25--0.5, diamonds) and CT ($>0.5$, open squares). The
B800 ring has only local states in this range. Twelve further B800 states at
3.1--3.3~eV, including nine local and three CT, are not shown.}
\label{fig:large_molecules}
\end{figure*}

NARF-mol, trained on $\omega$B97X-D only, also transfers to functionals
absent from training: the range-separated hybrid CAM-B3LYP and the global
hybrid PBE0. For CAM-B3LYP, the mean singlet RMSE is 36~meV, or 25~meV with
the $\Delta\Omega$ correction; for PBE0, where no correction is applied, it
is 15~meV. The S$_1$ oscillator strengths (Fig.~\ref{fig:transfer_chromophores}b) and
S$_1$--T$_1$ gaps (Fig.~\ref{fig:transfer_chromophores}c) are reproduced across
functionals as well. The one visible outlier, CBP with $\omega$B97X-D,
arises from state mixing within a nearly degenerate S$_1$/S$_3$ pair; with
the full AO basis and only the RI truncation, its S$_1$ oscillator strength
improves from 0.63 to 0.45 (reference 0.37). Overall, NARF's performance
across functionals suggests a shared rank structure among the response
kernels of different functionals, even across families (global and
range-separated hybrids).\par
To also reduce the ground-state cost, we introduce a simple scheme that yields
def2-SVPD spectra at the cost of a def2-SVP ground state. Diffuse functions may be needed to
describe delocalized excited states accurately, but they make the
ground-state SCF considerably more expensive. Because diffuse functions are
expected to change the ground-state density much less than the virtual
space~\cite{Papajak2011BasisSetsBeautiful}, we converge the ground state in the more compact def2-SVP basis
and update only the virtual orbitals, once, in the diffuse def2-SVPD basis,
before projecting them onto the truncated AO basis predicted by NARF. We pair this
one-shot update with the more compact NARF (\textit{sp}) and NARF
(\textit{sp/s}) settings, as we do below for LH2; they keep 53\% of the AO
basis and 7\% of the auxiliary basis (2\% for exchange with NARF
(\textit{sp/s})). For DMAC-TRZ and H$_2$TPP, the 30 singlet excitation
energies from this scheme deviate from the full def2-SVPD reference by RMSEs
of 28--29~meV and 29--30~meV, respectively, across the two settings, and the
brightest band shifts by at most 19~meV (Fig.~\ref{fig:transfer_chromophores}d,e).
Compared with calculations starting from a def2-SVPD ground state, the
one-shot update itself adds only 6--14~meV to the error (Fig.~S4).
The scheme is described in detail in Supplementary Note~8.

\begin{figure*}[!t]
    \centering
\includegraphics[width=1\linewidth]{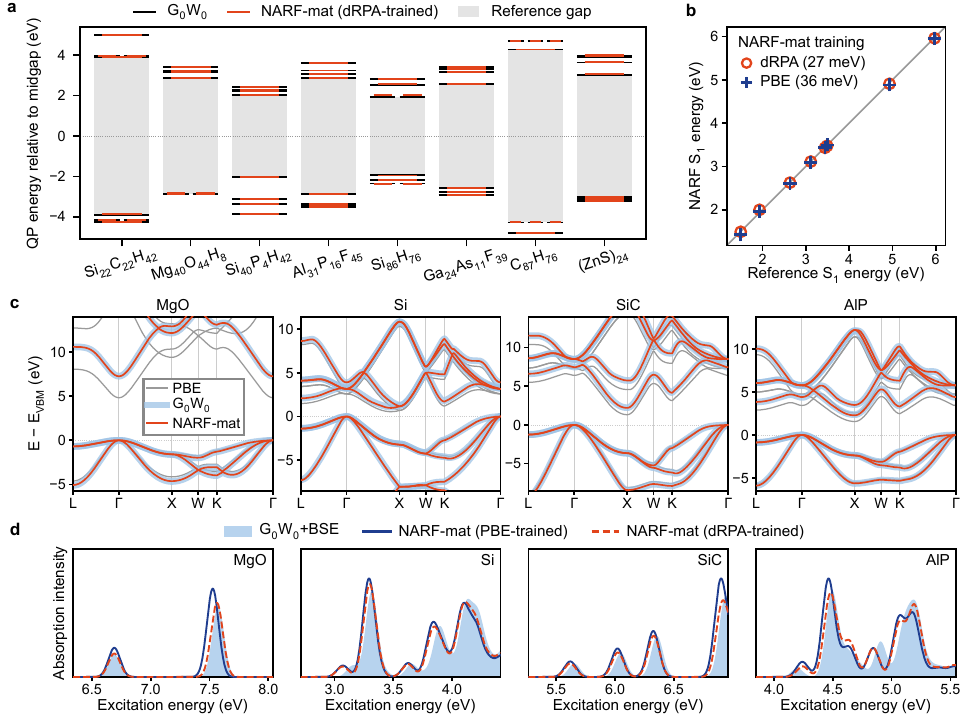}
\caption{\textbf{Transfer of NARF-mat to \textit{GW} and BSE in nanoclusters
and solids.}
NARF-mat truncates only the auxiliary basis and was trained on nanoclusters,
with no periodic data.
\textbf{a} Near-gap quasiparticle (QP) levels of the eight held-out clusters
from full-basis $G_0W_0$@PBE and from dRPA-trained NARF-mat with the
$\Delta\epsilon^{\mathrm{QP}}$ correction. Degenerate levels are
drawn side by side. Si$_{86}$H$_{76}$ contains a silicon vacancy, and
Si$_{40}$P$_4$H$_{42}$ is phosphorus-doped.
\textbf{b} Lowest $G_0W_0$+BSE@PBE excitation energy (S$_1$) for each cluster shown in
\textbf{a}, comparing NARF-mat with the full-basis reference. The auxiliary basis is
predicted by the model trained on the dRPA or the PBE kernel. RMSEs are given in
parentheses in the legend.
\textbf{c} Band structures of the periodic solids MgO, Si, SiC and AlP from
PBE, full-basis $G_0W_0$@PBE, and dRPA-trained NARF-mat. The NARF-mat $G_0W_0$@PBE
bands are computed without the $\Delta\epsilon^{\mathrm{QP}}$ correction. The bands are interpolated from
$8\times8\times8$ $k$-meshes and aligned to the valence band maximum (VBM).
\textbf{d} $G_0W_0$+BSE@PBE absorption spectra for the four solids shown in
\textbf{c}, computed within the TDA ($8\times8\times8$ $k$-mesh, Gaussian
broadening $\sigma = 0.05$~eV) from full-basis calculations and from the
PBE- and dRPA-trained NARF-mat models. Both BSE calculations use the
NARF-mat quasiparticle energies from \textbf{c}. The spectra include only
direct optical transitions.}
\label{fig:gw_bse}
\end{figure*}

\subsection{Large-scale application to LH2 pigment rings}
To demonstrate NARF on very large systems that challenge conventional
methods, we simulate the optical spectra of LH2 from \textit{Rhodoblastus
acidophilus}~\cite{cignoni2023machine}. We took the 27 bacteriochlorophyll \textit{a} (BChl
\textit{a}) pigments embedded in the protein core of Protein Data Bank entry
2FKW~\cite{cherezov2006room} as the photoactive core. They form two concentric rings,
which we treat separately: the 9 outer BChl \textit{a} molecules (B800) and
the 18 more tightly packed ones (B850) (Fig.~\ref{fig:large_molecules}a). We target
$\omega$B97X-D TDA excited states in the def2-SVPD basis, in which
$N_\mathrm{AO} = 17{,}379$ for B800 and $34{,}758$ for B850.

We converge the ground state in the def2-SVP basis. To reduce the cost of the one-shot
virtual-orbital update, we add diffuse functions on the pigment cores and
retain def2-SVP on the phytyl tails. The NARF AO basis contains 34\% of the full def2-SVPD basis, and TDA
is solved in an orbital window from 90~eV below the HOMO to 40~eV above the
LUMO. For the auxiliary basis we use NARF (\textit{sp/s}), the most compact
of the three settings, because at this size the exchange terms dominate the
cost and memory of the TDA solve; this combination was validated on DMAC-TRZ
and H$_2$TPP above. The RI basis retains 7\% of the full auxiliary basis for
the Coulomb and $f_\mathrm{xc}$ terms and 2\% for exchange. To further
reduce the cost of computing and storing integral intermediates, we use a
local-exchange and integral-packing strategy based on intrinsic atomic
orbitals~\cite{knizia2013intrinsic} (Supplementary Note~11). The reported energies include
the $\Delta\Omega$ corrections, which account for the basis truncation but
not for the local-exchange and local-$f_\mathrm{xc}$ approximations.

The B850 ring produces an intense band at 1.46~eV, formed by the strongly
coupled excitonic states of the tightly packed ring, while the
more weakly coupled B800 ring absorbs at 1.76~eV (Fig.~\ref{fig:large_molecules}b).
The calculations reproduce the experimental ordering and greater peak
intensity of B850 relative to B800. The calculated peaks lie about 0.01 and
0.21~eV above the respective room-temperature absorption maxima of 1.45 and
1.55~eV~\cite{georgakopoulou2002lh2}.
Charge-transfer states between
neighboring BChl \textit{a} units appear from 1.9~eV upward in B850 and
above 3.1~eV in B800 (Fig.~\ref{fig:large_molecules}c). The TDA solve took 0.9~h for
B800 (30 roots, two H100 GPUs) and about 20~h for B850 (60 roots in two
30-root windows, four H100 GPUs; Supplementary Note~10). These results show that NARF makes range-separated hybrid TDDFT practical for systems with thousands of atoms on a single GPU node.

\subsection{Transfer to \GW{} and BSE in nanoclusters and solids}
Predicting the quasiparticle and optical properties of functional materials
poses several practical challenges for approaches built on computational
basis sets. Nanoscale materials require large basis sets to cover all atoms
of a large nonperiodic system. Bulk materials need fewer basis functions,
but many $k$-points must be sampled to approach the thermodynamic limit,
which magnifies the cost of each basis function. Accurate simulations also
often require many-body theories with higher formal scaling than DFT. To
address these challenges, we apply NARF to \textit{GW} and BSE calculations
of inorganic nanoclusters and solids. We trained NARF-mat on 112
nanoclusters spanning 17 elements and 6--76 atoms, including hydrogen- and
fluorine-passivated clusters. Here NARF compresses only the RI auxiliary basis,
with an even-tempered auxiliary basis as the parent. For $G_0W_0$, the
covariates are computed from the static direct random phase approximation
(dRPA) kernel, and the BSE solver reuses an auxiliary basis learned from
either the dRPA or the PBE kernel, without any BSE-specific training. The
auxiliary bases learned from these static mean-field kernels are then used
in many-body \textit{GW} and BSE calculations.

On eight held-out clusters, including a Si vacancy, a P-doped Si cluster and
a (ZnS)$_{24}$ cage, the predicted auxiliary bases retain 13--31\% of the
parent functions. For each cluster, we averaged the absolute $G_0W_0$@PBE
quasiparticle (QP) energy errors over 42 orbitals, from HOMO$-$20 to LUMO$+$20. The QP
correction $\Delta\epsilon^{\mathrm{QP}}$ reduces the median of these
per-cluster errors from 37 to 4.3~meV. The corrected QP gaps have a mean
absolute error of 23~meV across the eight clusters relative to full-basis
$G_0W_0$@PBE (Fig.~\ref{fig:gw_bse}a). The silicon vacancy cluster is an exception: there the
correction increases the gap error from $-18$ to 92~meV, which may reflect the
absence of vacancy-type clusters from the training set. For comparison, placing a ghost atom at the vacancy site shifts the
full-basis HOMO and LUMO by $-16$ and $-34$~meV, which narrows the gap by
18~meV. The uncorrected NARF gap error of $-18$~meV is therefore of the same
order as this subtle basis-set effect. For BSE, the lowest excitation of the eight clusters agrees with full-basis
$G_0W_0$+BSE@PBE to an RMSE of 27~meV with the dRPA-trained model and 36~meV with the
PBE-trained model (Fig.~\ref{fig:gw_bse}b).

Without any training on periodic
systems, we then infer the auxiliary basis of bulk MgO, Si, SiC and AlP from
embedded clusters and solve periodic $G_0W_0$@PBE and
BSE~\cite{zhu2021allelectron,lei2022gaussianbased,Hillenbrand2025bse} on an
$8\times8\times8$ $k$-mesh. On average, 25\% (dRPA-trained) and 14\%
(PBE-trained) of the even-tempered auxiliary functions per primitive cell
are retained. The NARF-mat (dRPA-trained) bands follow the full-basis $G_0W_0$ bands
(Fig.~\ref{fig:gw_bse}c), with band-gap errors of $-8$, $-94$, $-29$ and
$-80$~meV for MgO, Si, SiC and AlP, respectively. These errors can be
controlled by the truncation: retaining 99\% of the cumulative weight in
the $s$, $p$ and $d$ shells together with all parent $f$ and $g$ functions
brings them below 1~meV for all four solids.
Since NARF-mat was trained only on $s$, $p$ and $d$
shells at a 90\% cumulative weight, we expect training with $f$ and $g$
shells and a higher weight to reduce the errors in solids. Finally, BSE
spectra computed on top of the NARF $G_0W_0$ quasiparticle energies
reproduce the full-basis line shapes, with S$_1$ errors between $-89$ and
$+45$~meV (Fig.~\ref{fig:gw_bse}d). These errors come almost entirely from the
quasiparticle energies: the truncated dRPA-trained auxiliary basis
contributes at most 7~meV through the BSE kernel. Better NARF quasiparticle
energies for solids would therefore directly improve the BSE spectra.

\section{Discussion}
We have introduced neural adaptive response functions (NARF) for learning
compact and interpretable representations of electron--hole interactions.
NARF combines atom-centered basis prediction with explicit response
calculations, providing access to excitation energies, oscillator strengths
and excited-state forces at substantially reduced cost. The response-kernel covariates
rank local basis modes by their contribution to the response, so the retained
space can be enlarged systematically. The transition-metal complexes
illustrate the value of this control. Retaining additional modes near the
metal improves the description of low-lying excitations with little increase
in the total auxiliary basis. A complementary $\Delta$-ML correction reduces residual
errors in excitation and quasiparticle energies relative to the full parent
basis.

The transferability of these representations offers insight into the basis
requirements of electronic response. NARF-mol, trained on a few hundred small
chromophores with one functional, transfers to larger molecules and to
functionals absent from training. A separate model, NARF-mat, learns
auxiliary bases from static mean-field kernels of nanoclusters and uses them
in $GW$ and BSE calculations of nanoclusters and periodic solids. These
results suggest that different response theories can use many of the same
local functions even when they couple electron--hole transitions differently.
The target calculation determines that coupling across the system, allowing
locally predicted bases to describe delocalized excitations and many-body
screening. Compact bases for expensive response methods can therefore be
learned from inexpensive reference calculations on small systems.

Together, these results show that NARF can greatly extend the scale of
excited-state calculations, as illustrated by range-separated hybrid TDDFT
spectra of a 2,520-atom LH2 pigment ring obtained on a single GPU node.
Several extensions follow naturally, including open-shell systems,
higher-angular-momentum auxiliary shells for solids, and an improved
treatment of condensed-phase environments. These developments could bring
accurate first-principles predictions of optical and quasiparticle properties
to a wider range of molecules and materials.

\section{Methods}
\subsection{Particle--hole response kernels}
In this work, we start from a single reference ground-state from either DFT or Hartree-Fock. Considering a spin-restricted closed-shell ground state for simplicity, the $N$-electron ground-state $\vert\Phi_0\rangle$ is a Slater determinant of doubly-occupied, single-particle states $\phi_p(r)$, (i.e. the molecular orbitals, MOs) with energy levels $\epsilon_p$. Each MO (indexed $p,q$) is expanded in a basis of atom-centered Gaussian functions $\chi_{\mu}(r)$ (indexed $\mu,\nu$). Going forward, we assume $\phi_p(r)$ and $\epsilon_p$ have already been obtained by solving the Kohn-Sham equations self-consistently \cite{kohn1965selfconsistent}. Then, to model the bound excited states of the $N$-electron system in second quantization, electrons are promoted from occupied states (indexed $i,j,k,l$) to virtual states (indexed $a,b,c,d$):

\begin{equation}
\vert{}\Phi_i^a\rangle = \hat{a}_a^\dagger \hat{a}_i \vert{}\Phi_0\rangle.
\end{equation}
Though the corresponding exciton Hamiltonian depends on the specific response theory, one often encounters a general form,
\begin{equation}
H_{ia, jb} = \langle \Phi_i^a \vert{} \hat{H} \vert{} \Phi_j^b \rangle = (\epsilon_a - \epsilon_i)\delta_{ij}\delta_{ab} + \langle ia |\mathcal{K} |jb\rangle,
\end{equation}
where $\epsilon_p$ are the single-particle energy levels and  $\mathcal{K}$ is the particle-hole interaction kernel, i.e. the response kernel. The exciton states $\vert{}\Phi^{(n)}\rangle$ are then expanded in the basis of hole removal plus particle addition,
\begin{subequations}
\begin{align}
    \vert{}\Phi^{(n)}\rangle = \sum_{i,a} Z^{(n)}_{ia}\vert{}\Phi_i^a\rangle = \sum_{i,a} Z^{(n)}_{ia} \, \hat{a}_a^\dagger \hat{a}_i \vert{}\Phi_0\rangle\\
    \sum_{jb} H_{ia, jb}Z^{(n)}_{jb} = \Omega^{(n)} Z^{(n)}_{ia},
\end{align}
\end{subequations}
 where the excitation energy and state vector of the $n$-th excited state are denoted $\Omega^{(n)}$ and $Z^{(n)}_{ia}$ respectively. Throughout this work, we invoke the commonly employed Tamm-Dancoff approximation (TDA), where the coupling between excitation and de-excitation amplitudes is neglected \cite{hirata1999timedependent}. In practice, $H_{ia, jb}$ is often too large to explicitly construct and diagonalize. Instead, matrix-free approaches such as the Davidson algorithm solve the low energy spectrum by repeated actions of the Hamiltonian on trial states $\vert{}\Phi^{(n)}\rangle$  \cite{Hillenbrand2025bse,stratmann1998efficient} \par
In TDDFT, expanding the response kernel in a basis of $ia$ pairs gives a matrix representation that depends primarily on the ERIs. In this work, we consider the $\omega$B97X-D functional, a range-separated hybrid (RSH) of the general form~\cite{chai2008longrange,chai2008systematic}:
\begin{subequations}
\begin{align}
\langle ia |\mathcal{K}^\text{TDDFT} |jb\rangle &= \kappa_1(ia \vert jb) - \kappa_2 (ij \vert ab) \nonumber\\
&\quad - \kappa_3 (ij \vert ab)_\gamma + \kappa_4 (ia|\delta^2 E_\text{xc}|jb)\\
(pq \vert rs) &= \iint \phi_p^*(\mathbf{r}_1) \phi_q(\mathbf{r}_1) \nonumber\\
&\quad\times \frac{1}{\vert\mathbf{r}_1 - \mathbf{r}_2\vert}\, \phi_r^*(\mathbf{r}_2) \phi_s(\mathbf{r}_2)\, d\mathbf{r}_1 d\mathbf{r}_2\\
(ij \vert ab)_{\gamma} &= \iint \phi_i^*(\mathbf{r}_1) \phi_j(\mathbf{r}_1) \nonumber\\
&\quad\times \frac{1 - \operatorname{erf}(\gamma \vert\mathbf{r}_1 - \mathbf{r}_2\vert)}{\vert\mathbf{r}_1 - \mathbf{r}_2\vert} \nonumber\\
&\quad\times \phi_a^*(\mathbf{r}_2) \phi_b(\mathbf{r}_2)\, d\mathbf{r}_1 d\mathbf{r}_2\\
(ia|\delta^2 E_\text{xc}|jb) &= \iint \phi_i^*(\mathbf{r}_1) \phi_a(\mathbf{r}_1) \nonumber\\
&\quad\times \left[ \frac{\delta^2 E_{\text{xc}}}{\delta n(\mathbf{r}_1) \delta n(\mathbf{r}_2)} \right] \nonumber\\
&\quad\times \phi_j^*(\mathbf{r}_2) \phi_b(\mathbf{r}_2)\, d\mathbf{r}_1 d\mathbf{r}_2.
\end{align}
\end{subequations}
The coulomb and exchange interactions come directly from the MO-space 4-index ERI $(pq|rs)$. The range separation parameter $\gamma$ controls the partitioning of the exchange integrals into short and long range components and $\kappa$ are fixed scalar parameters that depend on the specific functional and excited state spin multiplicity. For triplet states, the direct Coulomb term vanishes ($\kappa_1=0$). The density hessian $(ia|\delta^2 E_\text{xc}|jb)$ contains contributions from the exchange correlation energy $E_\mathrm{xc}$ that depend on the electron density and density gradients in real-space and cannot be expressed in terms of $(pq|rs)$. \par
Neglecting all terms in the TDDFT kernel besides the coulomb repulsion ($\kappa_1=1,\kappa_2=\kappa_3=\kappa_4=0$) gives us the direct RPA kernel 
\begin{equation}
\label{eq:drpa}
    \langle ia |\mathcal{K}^\text{dRPA} |jb\rangle = (ia \vert jb).
\end{equation}
This simple (static) interaction kernel is fundamental to $GW$ methods. We direct the readers to References~\citenum{zhu2021allelectron,lei2022gaussianbased,blase2020bethe} for methodological details for GTO-based \GW~ and BSE formalisms.
\subsection{Gaussian density fitting}
The dominant cost in applying the particle--hole interaction kernel $\mathcal{K}$ arises from contractions involving the four-index ERIs. To address this challenge, Gaussian density fitting introduces an auxiliary basis (indexed $P,Q$) to fit the pair-product density:
\begin{subequations}\label{eq:gdf}
\begin{align}
(pq|rs) &\approx \sum_{P,Q} (pq|P)\,[\mathbf V^{-1}]_{PQ}\,(Q|rs)\\
(pq|P) &= \iint \phi_p(\mathbf r_1)\phi_q(\mathbf r_1) \nonumber\\
&\quad\times \frac{1}{|\mathbf r_1-\mathbf r_2|}\,P(\mathbf r_2)\,d\mathbf r_1 d\mathbf r_2, \\
V_{PQ} &= (P|Q) \nonumber\\
&= \iint P(\mathbf r_1)\,\frac{1}{|\mathbf r_1-\mathbf r_2|}\,Q(\mathbf r_2)\,d\mathbf r_1 d\mathbf r_2 .
\end{align}
\end{subequations}
For each orbital pair $pq$, the fitting coefficients $d_{pq}^P$ minimize the Coulomb norm of the residual density $\delta \rho(\mathbf{r}) = \rho_{pq}(\mathbf{r}) - \sum_P d_{pq}^P P(\mathbf{r})$ \cite{dunlap1979approximations,vahtras1993integral}. Symmetric decomposition of $\mathbf V^{-1}$ allows us to evaluate Eq.~\ref{eq:gdf}a using a single set of three-center factors, $(pq|\bar P)=\sum_Q (pq|Q)[\mathbf V^{-1/2}]_{QP}$.
\par
This factorization reduces the cost of applying the TDDFT Coulomb kernel to $Z_{jb}$ from $O(N_\text{occ}^2N_\text{vir}^2)$ to $O(N_\text{RI}N_\text{occ}N_\text{vir})$. Density fitting also reduces integral storage by the same scalings, which allows one to store integrals in memory and avoid repeatedly recomputing integrals for each kernel action. In this work, we also devise a robust density-fitting scheme for $(ia|\delta^2 E_\text{xc}|jb)$, the term that arises from the generalized gradient approximation (GGA) contribution to $E_\text{xc}$, whose error is second order in the fitting residual and which incurs negligible accuracy loss. This approach avoids repeated real-space integrations and unifies our density fitting representations. Details can be found in Supplementary Note~9.

\subsection{Covariates for orbital and auxiliary bases}
\label{subsec:ARF}
We now introduce the central object of NARF, the response kernel covariate, which maps response kernel signal from particle-hole transitions onto atom-localized basis functions. NARF targets basis truncation at two independent stages: the primary AO basis for expanding the MOs and exciton states and the auxiliary basis for fitting orbital-pair products. First, to compact the primary AO basis, we derive single-particle covariates that generate a minimal set of adapted AOs from a large computational AO basis (e.g. used to calculate the ground-state). For atom $A$ and angular shell $\ell$, we calculate overlap integrals $O$ between AOs $(\chi_{A\ell m})_{\mu}$ and MOs and actions of the response kernel $\mathcal{K}$ on transition probes $Z^{(t)}$
\begin{subequations}
\begin{align}
(O_{A\ell m})_{\mu,p}=\langle(\chi_{A\ell m})_{\mu}\mid\phi_p\rangle\\
K_{ia}^{(t)}=
\sum_{jb}\langle ia |\mathcal{K} |jb\rangle\,Z^{(t)}_{jb}.
\end{align}
\end{subequations}
To prioritize ground-state density agreement, we define a covariate $M^0$ purely in terms of the ground-state occupied density overlap:
\begin{equation}
\label{eq:cov_ground}
(M_{A\ell}^0)_{\mu\nu}
=\sum_m\sum_{i\in \text{occ}}
(O_{A\ell m})_{\mu,i}\,
(O_{A\ell m})_{\nu,i}.
\end{equation}
Its leading overlap-orthogonal modes, as many as the MINAO count of the corresponding element and angular shell, are always retained (Eq.~\ref{eq:modesel} below). To capture response kernel signal missed by $M^0$, we define covariates for the hole channel $h$ and the particle channel $e$ by mapping the kernel actions $K^{(t)}_{ia}$ onto the overlaps 
\begin{subequations}
\begin{align}
(s^{(h,t)}_{A\ell m} )_{\mu a}&=
\sum_i (O_{A\ell m})_{\mu,i}K^{(t)}_{ia}\\
(s^{(e,t)}_{A\ell m} )_{\mu i}&=
\sum_a (O_{A\ell m})_{\mu,a}K^{(t)}_{ia}
\\
(M_{A\ell}^h)_{\mu\nu}
&=
\sum_{m,t,a}
(s^{(h,t)}_{A\ell m} )_{\mu a}(s^{(h,t)}_{A\ell m} )_{\nu a}
\\
(M_{A\ell}^e)_{\mu\nu}
&=
\sum_{m,t,i}
(s^{(e,t)}_{A\ell m} )_{\mu i}(s^{(e,t)}_{A\ell m} )_{\nu i}.
\end{align}
\end{subequations}
\par
The probes $Z^{(t)}$ are chosen so that the covariate receives signal from the span of particle-hole transitions relevant to the kernel response. For sufficiently small transition spaces (when $N_\text{occ}N_\text{vir}<5000$), $Z^{(t)}$ are chosen to be unit vectors spanning all possible $i\rightarrow a$ transitions. For larger transition spaces in the molecular TDA calculations, we sample five independent batches of $\lceil 0.2N_\text{occ}N_\text{vir}\rceil$ unit vectors each, without replacement within each batch, using Boltzmann weights $p_{ia}\propto\exp[-(\epsilon_a-\epsilon_i)/\tau_z]$ with $\tau_z=0.5$~eV. We average the five covariates before normalization. Diagonalizing $M_{A\ell}^0$ for all atom-shell pairs generates new AOs that best represent the ground-state density, while diagonalizing $M^{h}_{A\ell}$ and $M^{e}_{A\ell}$ reveals the modes that best represent the hole and particle channels of the response function (Eqs.~\ref{eq:orth} and \ref{eq:modesel} below). The hole and particle covariates $M^h$ and $M^e$ have units of energy squared and measure the response kernel signal mapped onto AO primitives over the transition probes.

\par In practice, the primary gains in NARF come from targeting the auxiliary density fitting basis for compression. We term the corresponding object the two-particle covariate. First, for each atom $A$ and angular-momentum shell $\ell$, we define a three-center coulomb integral between MO $p$, MO $q$, and auxiliary basis primitive $G_{A\ell m\mu}$. From these integrals, kernel actions
$K_{ia}^{(t)}$ are mapped onto $G_{A\ell m\mu}$ as 
\begin{subequations}
\begin{align}
\left(B_{A\ell m}\right)_{\mu,pq}
&=
\left(G_{A\ell m\mu}\middle|\phi_p\phi_q\right)_{1/r},
\\
\left(F^{(t)}_{A\ell m}\right)_{\mu}
&=
\sum_{ia}
\left(B_{A\ell m}\right)_{\mu,ia}
K^{(t)}_{ia}
\end{align}
\end{subequations}
where $\mu,\nu$ index a predefined set of auxiliary Gaussian primitives in each angular shell and $m$ indexes angular components of each shell. Summing the outer products over angular components $m$ and transition probes $t$ gives the auxiliary Coulomb covariate $M^J$
\begin{equation}
(M^J_{A\ell})_{\mu\nu}=\sum_m\sum_t
(F^{(t)}_{A\ell m})_\mu\,
(F^{(t)}_{A\ell m})_\nu.
\end{equation}
Analogous to the AO compression scheme, $M^J_{A\ell}$ is orthogonalized in the local Coulomb metric $V^{\mathrm{loc}}_{A\ell}$ of the auxiliaries in subshell $A\ell$ and diagonalized (Eqs.~\ref{eq:orth} and \ref{eq:modesel} below), which gives contractions over auxiliary primitives ranked by their contribution to the total kernel action signal. We also define a local exchange variant $M^{\text{x}}$, which we find helpful in certain cases where same-atom exchange is important:
\begin{subequations}
\label{eq:covx}
\begin{align}
T^{(t)}_{\mu m,\nu n} &= \frac12 \sum_{ijab} K^{(t)}_{ia}K^{(t)}_{jb} \nonumber\\
&\quad\times \left[ B_{\mu m,ij}B_{\nu n,ab} + B_{\nu n,ij}B_{\mu m,ab} \right] \\
M^{\text{x}}_{\mu\nu} &= \sum_{m}\sum_{t} \left[ \left|T^{(t)}\right|_V \right]_{\mu m,\nu m}
\end{align}
\end{subequations}
where $\left|T^{(t)}\right|_V$ denotes the spectral absolute value in $(V^{\mathrm{loc}}_{A\ell})$-orthogonal coordinates. Individual $T^{(t)}$ need not be positive semidefinite (PSD) in this construction, so Eq. \ref{eq:covx}b ensures large opposite sign contributions do not cancel between different probes. We direct the readers to Supplementary Note~12 for more details and an example system where same-atom exchange benefits from adding modes of $M^{\text{x}}$. 

\par
All covariates are turned into basis functions by the same procedure. Each covariate $M_{A\ell}$ ($M^0$, $M^h$ and $M^e$ in the primary basis; $M^J$ and $M^{\text{x}}$ in the auxiliary basis) is first symmetrically orthogonalized in its local metric and normalized to unit trace,
\begin{subequations}
\label{eq:orth}
\begin{align}
\hat M^{0,h,e}_{A\ell} &= \frac{(S^{\mathrm{loc}}_{A\ell})^{-1/2}\,M^{0,h,e}_{A\ell}\,(S^{\mathrm{loc}}_{A\ell})^{-1/2}}{\operatorname{Tr}[(S^{\mathrm{loc}}_{A\ell})^{-1}M^{0,h,e}_{A\ell}]}\\
\hat M^{J,\text{x}}_{A\ell} &= \frac{(V^{\mathrm{loc}}_{A\ell})^{-1/2}\,M^{J,\text{x}}_{A\ell}\,(V^{\mathrm{loc}}_{A\ell})^{-1/2}}{\operatorname{Tr}[(V^{\mathrm{loc}}_{A\ell})^{-1}M^{J,\text{x}}_{A\ell}]}
\end{align}
\end{subequations}
where $(S^{\mathrm{loc}}_{A\ell})_{\mu\nu}$ is the local overlap between the AO primitives $\mu,\nu$ of subshell $A\ell$ and $(V^{\mathrm{loc}}_{A\ell})_{\mu\nu}$ the local Coulomb metric between its auxiliary primitives. This removes the arbitrary energy scale of the kernel actions and the metric dependence of the primitives. The neural network learns the normalized $M^h$, $M^e$, $M^J$, and $M^{\text{x}}$ covariates. We compute $M^0$ directly from the occupied orbitals.

\subsection{Mode selection and basis assembly}
\label{subsection:ABA}
The orthogonalized covariate is then diagonalized,
\begin{subequations}
\label{eq:modesel}
\begin{align}
\hat M_{A\ell}\,\tilde c^{(\alpha)} &= \lambda_\alpha\,\tilde c^{(\alpha)},\\
c^{(\alpha)} &= (G^{\mathrm{loc}}_{A\ell})^{-1/2}\,\tilde c^{(\alpha)},
\end{align}
\end{subequations}
where $G^{\mathrm{loc}}_{A\ell}$ stands for $S^{\mathrm{loc}}_{A\ell}$ or $V^{\mathrm{loc}}_{A\ell}$ and $c^{(\alpha)}$ are the resulting contractions over the primitives; the generalized eigenproblem for $M_{A\ell}$ and $G^{\mathrm{loc}}_{A\ell}$ gives the same contractions, with eigenvalues proportional to $\lambda_\alpha$. The modes are variationally optimal: the leading mode maximizes $c^{\mathsf T}M_{A\ell}c$ subject to $c^{\mathsf T}G^{\mathrm{loc}}_{A\ell}c=1$, and each subsequent mode maximizes the same quantity while remaining metric-orthogonal to the preceding ones, so for $M^0$ they are the contractions that best capture the ground-state density and for the other covariates those that best capture the kernel signal of the respective channel. Because $\sum_\alpha\lambda_\alpha=1$, each $\lambda_\alpha$ gives the fraction of the signal carried by mode $\alpha$. We define the effective rank of each atom and shell as
\begin{equation}
\label{eq:effrank}
r_{A\ell} = \frac{1}{\sum_\alpha \lambda_\alpha^2}.
\end{equation}
If $k$ modes share the signal equally and the remaining modes carry zero weight, $r_{A\ell}=k$.

For the AO basis, we keep as many leading $M^0$ modes as the MINAO count of the element and shell. We project $M^h$ and $M^e$ outside this retained space and add their residual modes with $\lambda_\alpha>\delta_\lambda$ (0.2 unless stated otherwise). For each retained non-hydrogen shell in the default molecular auxiliary basis, we keep the leading $M^J$ mode and any additional modes above $\delta_\lambda$, then apply the same threshold to $M^{\text{x}}$ projected outside the retained $M^J$ space. In both bases, residual eigenvalues are measured relative to the original covariate trace. Alternatively, a cumulative capture $\theta$ keeps the smallest number $k$ of leading modes with $\sum_{\alpha\le k}\lambda_\alpha\ge\theta$; unlike the per-mode threshold, this admits more modes when the signal is spread over several of them.
\par
We emphasize that AO compression is completely independent of auxiliary basis compression in our adaptive response framework. They may be used separately, together, or in conjunction with MO active space truncation, which highlights the flexibility of this approach.

To decide whether to add the next angular shell on atom $A$ to the primary AO basis, we introduce two importance descriptors. In particular, the more expensive $d$ shells (5 basis functions per radial contraction) can be reserved for only a few atoms. We derive these descriptors from ground-state information only, to avoid forming the response kernel in the larger basis. In terms of projections $\Pi$ onto two sets of atomic basis functions $\mathcal P$ and $\mathcal Q$ such that $\mathcal Q \subseteq \mathcal P$ these descriptors $f$ are given by
\begin{subequations}
\begin{align}
f_{A}^o
=
2\sum_{i\in\text{occ}}
\langle\phi_i\lvert\left(\Pi_{\mathcal P}-\Pi_{\mathcal Q}\right)
\lvert\phi_i\rangle\\
f_{A}^v=\sum_{a \in \text{vir}} p_a\langle\phi_a\lvert\left(\Pi_{\mathcal P}-\Pi_{\mathcal Q}\right)
\lvert\phi_a\rangle
\end{align}
\end{subequations}
where $p_a\propto e^{-(\varepsilon_a-\varepsilon_\text{LUMO})/\tau}$, normalized so that $\sum_a p_a=1$, with $\tau=0.5$ eV. Here $\mathcal Q$ spans all primitives in the MINAO-allowed angular shells on all atoms. The augmented space $\mathcal P$ adds the primitives of the candidate shell on atom $A$ alone, for example the $d$ shell on a carbon atom. Then, an atom is augmented with extra $d$ shells when $f^o_A>\delta_o$ or $f^v_A>\delta_v$ for tunable thresholds $\delta_o$ and $\delta_v$. For both spin channels, we use a general eigenvalue threshold of $\delta_\lambda=0.2$ for both auxiliary functions and AOs. To decide whether AOs on atom $A$ receive extra angular capacity, we set $\delta_o=0.01$ and $\delta_v=0.005$. Benchmarks that deviate from these defaults, namely the per-mode threshold of 0.1 and the 0.95 cumulative capture on transition metals and their coordinating atoms (Fig.~\ref{fig:tm}), the NARF~($sp$) and NARF~($sp/s$) auxiliary settings and the large-system reductions (Figs.~\ref{fig:timing}a,b, \ref{fig:transfer_chromophores}d,e and \ref{fig:large_molecules}), and the 0.90 cumulative capture used for nanocluster and periodic $G_0W_0$ calculations and the separate test for solids using 0.99 capture of the $s$, $p$ and $d$ modes plus all parent $f$ and $g$ functions (Fig.~\ref{fig:gw_bse}), are listed benchmark by benchmark in Supplementary Table~1 and Supplementary Notes~3--6.\par

We obtain the molecular-orbital coefficients $\widetilde C$ in the retained AOs $u_\kappa$ by projecting the coefficients $C$ from the parent AOs $\chi_\mu$,
\begin{equation}
\label{eq:ao_projection}
\widetilde C = (S^{u})^{-1} S^{u\chi} C,
\end{equation}
where $S^{u}_{\kappa\lambda}=\langle u_\kappa\mid u_\lambda\rangle$ and $S^{u\chi}_{\kappa\mu}=\langle u_\kappa\mid\chi_\mu\rangle$. This projection retains the input orbital energies without re-diagonalizing the ground-state Hamiltonian.

\subsection{\texorpdfstring{$\Delta$}{Δ}-ML corrections for basis truncation}
Finally, we present an approach for predicting energy errors on individual excited states. Following Fig.~\ref{fig:narf_scheme}a, we define the correction as $\Delta\Omega^n=\Omega^n_{\mathrm{full}}-\Omega^n_{\mathrm{trunc}}$ and add its predicted value to the truncated-basis excitation energy. We take inspiration from existing basis set correction schemes developed for post-DFT methods \cite{dutta2018NearlinearScalingEquation,duttaDomainbasedLocalPair2019,giner2019bsc,loos2020bsc} and data-driven quantum chemistry~\cite{ramakrishnan2015delta,welborn2018transfer}. Targeting energy corrections for the TDA excitation energies, we construct the following Löwdin descriptors $\mathbf{g}_{n,A}$ from the TDA eigenvectors $Z^{n}_{ia}$ and MO coefficients $C_{\mu p}$:
\begin{subequations}
\label{eq:dw_ao_features}
\begin{align}
f_A^{(e,h)} &= 1-\frac{1}{\ell_\text{max}+1}
  \sum_{\ell\in A} \mathrm{Tr}\big[ c^{T}_{\mathcal{Q}\ell} \hat{M}^{(e,h)}_{\mathcal{P}\ell} c_{\mathcal{Q}\ell} \big]\\
q^{h}_{n,A} &= \sum_{\mu \in A} \sum_{a}
    \Bigg( \sum_{\kappa i} S^{1/2}_{\mu\kappa} C_{\kappa i} Z^{n}_{ia} \Bigg)^{2},
    \label{eq:qh} \\[4pt]
  q^{e}_{n,A} &= \sum_{\mu \in A} \sum_{i}
    \Bigg( \sum_{\lambda a} S^{1/2}_{\mu\lambda} C_{\lambda a} Z^{n}_{ia} \Bigg)^{2},
    \label{eq:qe} \\[4pt]
\mathbf g_{n,A}&=
\Big(\;f^{h}_A\,q^{h}_{n,A},\;\; f^{e}_A\,q^{e}_{n,A},\;\; f^{o}_A\,q^{h}_{n,A},\;\; f^{v}_A\,q^{e}_{n,A}\;\;\Big).
\end{align}
\end{subequations}
$c_\mathcal{Q\ell}$ are the retained covariate modes of the full $\hat{M}^{(e,h)}$. We also define an analogous set of descriptors $\mathbf{G}_{n,A}$ that capture missing pair product density due to auxiliary basis truncation.
\begin{subequations}
\label{eq:dw_ri_features}
\begin{align}
f^{J}_{A\ell} &= 1-\operatorname{Tr}\!\big[c_{\mathcal Q\ell}^{\mathsf T}\,\hat M^{J}_{\mathcal P\ell}\,c_{\mathcal Q\ell}\big]\\
T^{n}_{\mu\nu} &= \sum_{\kappa\lambda} \sum_{ia}
    S^{1/2}_{\mu\kappa} C_{\kappa i}\, Z^{n}_{ia}\, C_{\lambda a} S^{1/2}_{\lambda\nu},\label{eq:Tuv} \\[4pt]
q^{t}_{n,A} &= \frac{1}{2} \sum_{\mu \in A} \sum_{\nu}
    \Big[ \big( T^{n}_{\mu\nu} \big)^{2} + \big( T^{n}_{\nu\mu} \big)^{2} \Big],
    \label{eq:qt} \\[4pt]
\mathbf{G}_{n,A\ell} &= \Big(\;f^J_{A\ell}\,q^{t}_{n,A},\;\; f^J_{A\ell}\,\bar q_{n,A},\;\; J^\mathrm{miss}_{n,A\ell}\;\Big), \\[4pt]
\bar q_{n,A} &= \tfrac12\big(q^{h}_{n,A}+q^{e}_{n,A}\big).
\end{align}
\end{subequations}
Here, $J^\mathrm{miss}_{n,A\ell}$ is the portion of the Coulomb kernel missed by the auxiliary truncation, localized to the neighborhood $\mathcal{N}(A)$ of atoms within 8~bohr of atom $A$. For the truncated-basis state, let $D^n=C_o Z^n C_v^{\mathsf T}$. This localized Coulomb element is obtained from the difference between its unfitted transition density $\rho^t_n(\mathbf r)$ and its fit in the retained auxiliary space. In the following expressions, $V_{\mathcal N(A)}^{+}$ denotes the pseudoinverse of the local Coulomb metric and $\tilde P$ denotes a retained auxiliary function.
\begin{subequations}\label{eq:jmiss}
\begin{align}
\rho^t_n(\mathbf r) &= \sum_{\mu\nu} D^n_{\mu\nu}\,\chi_\mu(\mathbf r)\chi_\nu(\mathbf r) \\[4pt]
\tilde\rho^t_n(\mathbf r) &= \sum_{\tilde P\tilde Q\in\mathcal N(A)} \tilde P(\mathbf r)\,\big[\tilde V^{-1}\big]_{\tilde P\tilde Q}\,(\tilde Q|\rho^t_n) \\[4pt]
J^{\mathrm{miss}}_{n,A\ell} &= \sum_{R\in(A,\ell)} \Big( \sum_{S\in\mathcal N(A)} \big[(V_{\mathcal N(A)}^{+})^{1/2}\big]_{RS} \nonumber\\
&\qquad\qquad\times (S|\rho^t_n - \tilde\rho^t_n) \Big)^2.
\end{align}
\end{subequations}
For correcting single quasiparticle excitations from NARF \GW~ calculations, we define a simplified descriptor set indexed by MO $p$:
\begin{subequations}\label{eq:gw_dqp}
\begin{align}
q_{p,A} &= \sum_{\mu \in A}
    \Bigg( \sum_{\kappa } S^{1/2}_{\mu\kappa} C_{\kappa p}\Bigg)^{2},
    \label{eq:qp} \\[4pt]
\mathbf g_{i,A} &= \Big(\;f^{h}_A\,q_{i,A},\;\;f^{o}_A\,q_{i,A}\;\Big), \\[4pt]
\mathbf g_{a,A} &= \Big(\;f^{e}_A\,q_{a,A},\;\;f^{v}_A\,q_{a,A}\;\Big), \\[4pt]
\mathbf{G}_{p,A\ell} &= \Big(\;f^J_{A\ell}\,q_{p,A},\;\; f^J_{A\ell}\,J^{\mathrm{miss}}_{p,A\ell}\;\Big)
\end{align}
\end{subequations}
where $J^\text{miss}_{p,A\ell}$ depends on the difference in full-space and truncated-basis MO densities in real-space $\phi^2_p(\mathbf{r}) -\tilde{\phi}^2_p(\mathbf{r})$.
\par
Finally, equivariant neural networks learn the state-resolved excitation energy corrections as a linear function of descriptors $\mathbf{g}$ and $\mathbf{G}$. The linear weights for AO truncation correction $\mathbf{w}_{A}$ and for auxiliary truncation correction $\mathbf{W}_{A\ell}$ are read out from the atom-wise embeddings of the neural networks as
\begin{subequations}
\begin{align}
\mathbf{w}_{A} &= D^\mathbf{w}(h^\mathbf{w}_A(\{Z_B,\mathbf{r}_B\}_{B\in d}))\\
\mathbf{W}_{A\ell} &= D^\mathbf{W}(h^\mathbf{W}_A(\{Z_B,\mathbf{r}_B\}_{B\in d}))\\
\Delta\Omega^n_\mathrm{AO} &=  \sum_{A}(\mathbf{w}_{A})^T\mathbf{g}_{n,A}\\
\Delta\Omega^n_\mathrm{RI} &=  \sum_{A\ell}(\mathbf{W}_{A\ell})^T\mathbf{G}_{n,A\ell}.
\end{align}
\end{subequations}
Note that though the descriptors $\mathbf{g}_{n,A}$ and $\mathbf{G}_{n,A\ell}$ depend on converged NARF TDA excited states and local Löwdin descriptors, the corresponding linear weights only depend on the atomic structure. In this work, we train a $\Delta\Omega^n_\mathrm{AO}$ model to predict corrections for NARF calculations with extra angular admittance tolerances $\delta_o=0.01$ and $\delta_v=0.005$ and a universal covariate capture threshold of $\delta_\lambda=0.2$. For each spin channel, we train a base RI correction model using auxiliary spaces with $\ell_{\max}=1$ and 2. We use this model for NARF (\textit{spd}) and fine-tune it separately for NARF (\textit{sp}) and NARF (\textit{sp/s}). Though the training data labels isolate one error source at a time ($\Delta\Omega^n_\mathrm{AO}$ or $\Delta\Omega^n_\mathrm{RI}$ relative to the full basis reference), inference may require a correction from both truncations simultaneously. In this case, we approximate the combined truncation error as $\Delta\Omega^n_\mathrm{AO,RI} \approx\Delta\Omega^n_\mathrm{AO}+\Delta\Omega^n_\mathrm{RI}$. The above $\Delta$-ML strategy is mirrored for learning the corrections to \GW~ quasiparticle energies $\Delta\epsilon^\text{QP}_p$.

\subsection{Neural network architecture and training}
\label{subsection:training}
All covariate, $\Delta\Omega$ and $\Delta\epsilon^\text{QP}$ models share one architecture: a GotenNet trunk~\cite{aykent2025gotennet} with three interaction layers, a feature width of 128 and a 5~\AA{} cosine cutoff, which takes only atomic numbers and Cartesian coordinates, followed by one readout head per target and spin channel. For each target and spin channel, the readout $D_\ell$ maps the embedding $\mathbf h_A$ of atom $A$ to a factor $F_{A\ell}=D_\ell(\mathbf h_A)$ for shell $\ell$. For each nonzero block, the predicted covariate used for mode selection is
\begin{equation}
\label{eq:pred_covariate}
\widehat M_{\mathrm{pred}}^{A\ell}
= \frac{F_{A\ell}F_{A\ell}^{\mathsf T}}
       {\operatorname{Tr}(F_{A\ell}F_{A\ell}^{\mathsf T})}.
\end{equation}
The factorization ensures positive semidefiniteness, and division by the trace gives unit total mode weight. Each head is trained with a masked Frobenius loss on the whitened, unit-trace covariates in which every molecule, target, spin channel and shell carries equal weight. NARF-mol was fine-tuned from a model trained from scratch on 698 molecules; NARF-mol-tm was fine-tuned from NARF-mol on its 706 molecules plus 208 tmQM complexes, with the two-particle head extended to $f$ and $g$ shells; NARF-mat was trained from scratch on 112 clusters. The covariate models are final fits on all of their training data and are evaluated on the held-out test sets described in the Results; for NARF-mat we additionally report five-fold cross-validation in which near-duplicate clusters share a fold. The $\Delta\Omega$ and $\Delta\epsilon^\text{QP}$ models are trained on the per-root (per-orbital) energy differences between full-basis and truncated calculations, with a mean-squared-error loss, five-fold molecule-level cross-validation and early stopping on an inner 10\% split, except that the transition-metal $\Delta\Omega$ models were refit on 914 molecules for a fixed number of epochs set in advance; the $\Delta\epsilon^\text{QP}$ prediction is the mean of a five-seed ensemble. The graph settings and hyperparameters of every model are listed in Supplementary Note~13.

\subsection{Computational details}
Unless otherwise specified, molecular TDA calculations use the $\omega$B97X-D functional, def2-SVPD as the parent AO basis, and our GPU implementation of density-fitted TDA. The full-basis references use the def2-universal-jkfit auxiliary basis for the Coulomb, exchange and $f_\mathrm{xc}$ terms, with $f_\mathrm{xc}$ fitted as in Supplementary Note~9, so that the reported NARF errors isolate the basis truncation; GPU4PySCF~\cite{li2025gpu4pyscf,wu2025gpu4pyscf} with grid-evaluated $f_\mathrm{xc}$ serves as the timing comparator in Fig.~\ref{fig:timing} and as the reference for the BChl monomer and dimer tests.

For Fig.~\ref{fig:small_molecules_results}c, we assigned the character
of the ten lowest singlet states of each test molecule using the full-basis
reference calculation. We classified a state as Rydberg if at least 50\%
of its excited-electron population lay on the diffuse functions of
def2-SVPD. For the remaining states, we computed the earth mover's
distance between the atom-resolved hole and electron populations,
adapting the charge-transfer metric of Wang et al.~\cite{wang2023emd}.
We classified states with a distance of at least 2.0~\AA{} as
charge-transfer states and all others as valence states.

Functional-transfer tests use CAM-B3LYP and PBE0 in the same basis. Geometries of the 17 large chromophores were optimized with ORCA~\cite{Neese2025ORCA6} at the $\omega$B97X-3c level~\cite{Muller2023wB97X3c}. Timings in Fig.~\ref{fig:timing}a,b were measured on a single NVIDIA H200 GPU; the LH2 calculations in Fig.~\ref{fig:large_molecules} ran on two or four H100 GPUs of one node with the local-exchange implementation of Supplementary Note~11. The B800 and B850 rings were extracted from a hydrogenated structure derived from PDB entry 2FKW and treated separately as neutral singlets. The Cartesian coordinates used for both calculations are provided in the manuscript repository.
The nanocluster and periodic PBE and $G_0W_0$+BSE calculations use the ccECP pseudopotentials~\cite{Bennett2017ccECP,Bennett2018ccECP,Wang2019ccECP} with the matching ccECP-cc-pVDZ basis, as distributed with ccgto~\cite{ye2022correlationconsistent}, for Be--S (H all-electron in the same basis), and the cc-pVDZ-PP basis~\cite{Peterson2003ccpvnzpp1,Peterson2003ccpvnzpp2,Peterson2005ccpvnzpp} with the 10-electron-core Stuttgart pseudopotentials (ECP10MDF)~\cite{Metz2000ecp,Figgen2005ecp} for Zn, Ga, Ge, As and Se. Auxiliary bases are even-tempered (aug-ETB, $\beta=2.0$)~\cite{sun2020recent}. We use the molecular $G_0W_0$ and BSE solvers in NARF for nanoclusters and the periodic $G_0W_0$ and BSE solvers in \texttt{fcdmft}~\cite{zhugroupyale2026fcdmft} for crystals. Quasiparticle energies are obtained by analytic continuation (Pad\'e, 18 points) of the self-energy evaluated on 100 imaginary-frequency points, with the quasiparticle equation solved by Newton iteration; the crystals use a $\Gamma$-centred $8\times8\times8$ $k$-mesh with a finite-size correction, and the periodic BSE is solved at $Q=0$ in the TDA with all occupied and the lowest seven virtual bands. For the periodic covariate predictions, the cell is embedded in a cluster of its periodic images beyond the receptive field of the network (Supplementary Note~6).

\backmatter

\bmhead{Data availability}
The data supporting the findings of this study will be made publicly available upon publication.

\bmhead{Code availability}
The NARF code used in this work will be made publicly available upon publication.

\bibliography{ref}

\bmhead{Acknowledgements}
This work was supported by a Cottrell Scholar Award (CS-CSA-226-036) and an AI seed grant from the Yale Dean of Science. C.V. acknowledges partial support from the Department of Defense through the National Defense Science and Engineering Graduate (NDSEG) Fellowship Program. The development of periodic \GW+BSE code was supported by National Science Foundation under Grant No.~CHE-2337991. This work used Expanse at the San Diego Supercomputer Center through allocation CHE260064 from the Advanced Cyberinfrastructure Coordination Ecosystem: Services \& Support (ACCESS) program, which is supported by U.S. National Science Foundation grants \#2138259, \#2138286, \#2138307, \#2137603, and \#2138296. We thank the Yale Center for Research Computing for guidance and use of the research computing infrastructure.

\bmhead{Author contributions}
C.V. conceived and designed the method, wrote the NARF code, performed the calculations, and wrote the original draft of the manuscript. T.Z. supervised the research and guided the project design. D.K. and K.L. contributed to methodological discussions and validation of the NARF implementation. J.L. performed periodic \GW+BSE calculations.  All authors discussed the results and contributed to writing and editing the manuscript.

\bmhead{Competing interests}
The authors declare no competing interests.

\bmhead{AI use declaration}
We used ChatGPT and Codex (GPT 6-6.2 Sol and 6-6.2 Astra) and Claude (Opus 5.5 and Fable 5.2) to assist with code editing, calculations, and revisions to the manuscript. We verified the code and results and take full responsibility for the content of this work.

\end{document}